\documentclass[11pt]{article}

\usepackage[preprint]{acl}

\usepackage{times}
\usepackage{latexsym}
\usepackage{amsfonts}

\usepackage[T1]{fontenc}

\usepackage[utf8]{inputenc}

\usepackage{microtype}

\usepackage{inconsolata}

\usepackage{graphicx}

\usepackage[breakable,skins]{tcolorbox}
\usepackage{enumitem}
\usepackage{fvextra}
\usepackage{multirow}
\usepackage{booktabs}
\usepackage{amsmath}
\usepackage[table]{xcolor}
\usepackage{xurl}      
\usepackage{hyperref}

\newtheorem{definition}{Definition}

\title{Enhancing the Socioeconomic Understanding of Foundation Models with Urban Mobility}

\author{
\textbf{Baoshen Guo\textsuperscript{1}},
\textbf{Donghang Li\textsuperscript{5}},
\textbf{Zhiqing Hong\textsuperscript{2}},
\textbf{Kailai Sun\textsuperscript{1}}, \\
\textbf{Heye Huang\textsuperscript{3}}, 
\textbf{Alok Prakash\textsuperscript{1}}, 
\textbf{Shenhao Wang\textsuperscript{4}}
\\
\textsuperscript{1}Singapore-MIT Alliance for Research and Technology (SMART), Singapore, \\
\textsuperscript{2}Hong Kong University of Science \& Technology (Guangzhou), China, \\
\textsuperscript{3}Korea Advanced Institute of Science \& Technology, Korea \\
\textsuperscript{4}University of Florida, Gainesville, USA, 
\textsuperscript{5}Massachusetts Institute of Technology, Cambridge, USA
 \\
}

\begin{document}
\maketitle

\begin{abstract}

Foundation models have recently been applied to urban socioeconomic prediction using POI text, satellite imagery, and geospatial descriptions. 
However, these models mostly rely on static attributes of individual places, while ignoring the mobility patterns that reveal how places are functionally connected. 
To address this gap, we explore whether mobility networks can elicit the geospatial capabilities of foundation models by explicitly encoding connectivity among urban entities. 
We propose \textit{MobFusion}, a modular mobility-enhanced foundation model fusion paradigm, and instantiate it through three complementary designs: (i) mobility networks as contexts for zero-shot LLM prompting, (ii) as graph connectors for fusing geospatial visual embeddings with textual embeddings, and (iii) as structured tokens for multimodal LLM reasoning. 
Using anonymized large-scale mobility datasets from three U.S. metropolitan areas, we find that \textit{MobFusion} improves urban prediction tasks (e.g., median household income, population density, and crime prediction) across three instantiations, demonstrating that incorporating human mobility can effectively improve the socioeconomic understanding of foundation models.
\end{abstract}

\section{Introduction}

Cities are complex relational systems composed of heterogeneous entities and intertwined interactions, and they can be described from multiple perspectives, including text~\cite{reades2025city,guo2025language}, imagery~\cite{quintana2025global,fan2023urban} such as satellite and street-view images, graph-structured relations~\cite{louail2015uncovering,alves2021commuting} such as social and mobility networks, and numerical measurements such as socioeconomic indicators. 
Recent advances in foundation models have achieved remarkable progress in representing the multimodal characteristics of cities. GeoLLM~\cite{manvi2023geollm,manvi2024large} leverages textualized geographic contexts to infer regional socioeconomic conditions, earth-observation foundation models such as AlphaEarth~\cite{brown2025alphaearth} provide globally scalable embeddings of the physical environment, and urban vision-language models integrate satellite, street-view, and textual information for socioeconomic sensing and prediction~\cite{liucitylens,liu2025cityrise,hao2025urbanvlp}.
However, these studies primarily characterize urban entities or regions through their intrinsic textual and visual attributes, with limited attention to the relational mobility context that reveals how places are functionally connected across the city.

Human mobility provides relational signals that geographic attributes alone cannot capture. Urban phenomena such as ghost cities~\cite{jin2017evaluating} and post-COVID changes~\cite{chang2021mobility} are often more clearly reflected in mobility patterns than in the built environment itself, highlighting the importance of movement-based information for understanding cities. 
Recent studies have incorporated human mobility into deep learning-based urban modeling, either by using graph neural networks to capture mobility relations for task-specific prediction~\cite{zhou2023heterogeneous,zhang2021multi,hui2020predicting} or by treating mobility as a modality for alignment with POI semantics and visual features~\cite{wen2026mora}.
However, systematically integrating human mobility across diverse foundation models remains challenging. For example, it remains unclear how to condense complex mobility networks into prompts for efficient zero-shot LLM reasoning, or how to inject mobility-derived relational structure into vision foundation models and MLLMs that primarily reason over visual and textual attributes of individual urban entities.

\begin{figure*}
    \centering
    \includegraphics[width=0.75\linewidth]{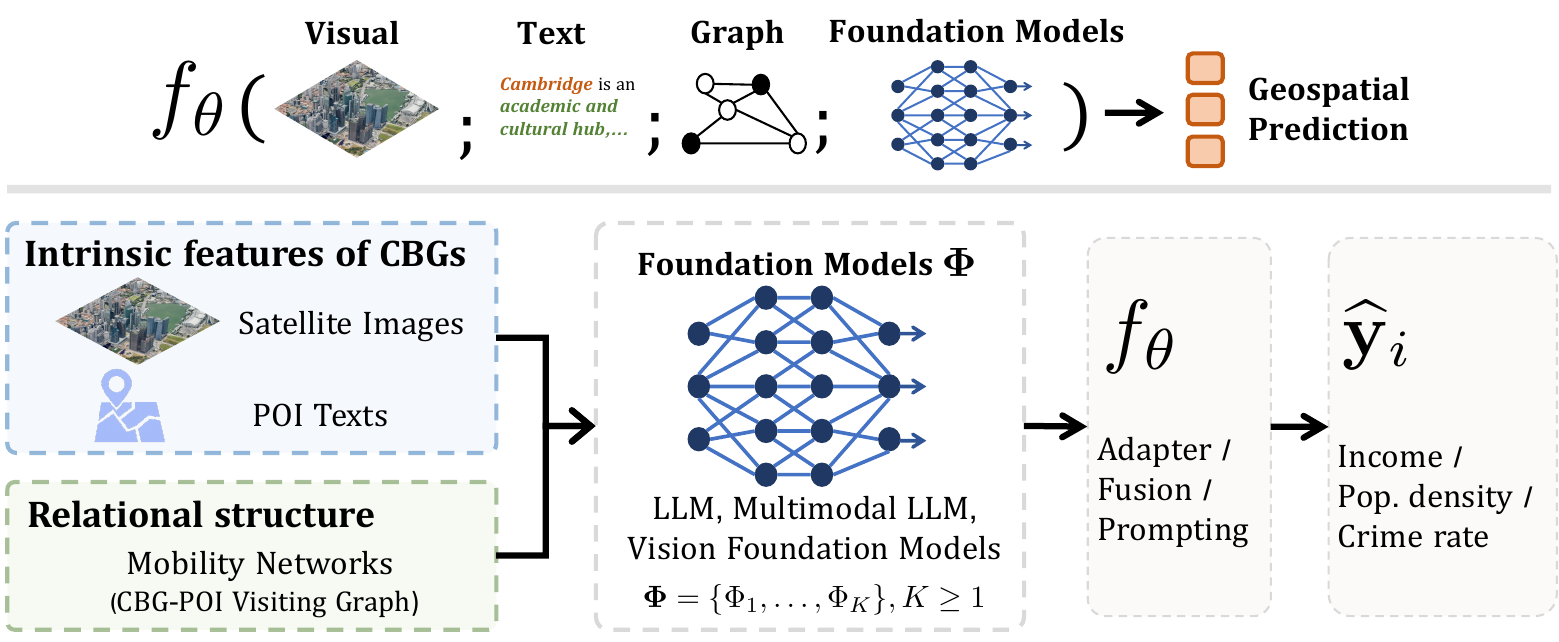}
    \caption{Framework of \textit{MobFusion}. Intrinsic census block group (CBG) features (Vision, POI text) and relational mobility network are encoded by a set of foundation models $\Phi$ and fused via $f_\theta$ for downstream geospatial prediction.}
    \label{fig:01-framework}
\end{figure*}

In this paper, we explore whether human mobility networks can serve as relational grounding signals that improve foundation models' ability to reason about urban socioeconomic conditions. 
Our premise is that foundation models capture rich textual and visual knowledge of urban entities, while human mobility provides complementary relational signals about how these entities are connected. 
Specifically, as shown in Figure~\ref{fig:01-framework}, we propose \textit{MobFusion}, a mobility-enhanced foundation model fusion framework grounded in modular AI and urban sociological insights. 
\textit{MobFusion} incorporates mobility in three forms: (i) compressed into textual contexts for zero-shot LLM prompting guided by mobility-based neighborhood effects~\cite{levy2020triple}; (ii) used as relational connectors for fusing earth-observation and textual descriptions of points of interest (POI); and (iii) mapped into graph tokens for multimodal LLM reasoning.
The contributions are as follows:
\begin{itemize}[leftmargin=*]
\item We propose a modular mobility-enhanced foundation-modeling paradigm for urban socioeconomic understanding, combining foundation models' strength in representing intrinsic geospatial attributes with mobility networks' ability to capture relational urban knowledge.

\item We construct the mobility networks between Census Block Groups (CBGs) and POIs, and develop three integration strategies: mobility as prompting contexts for LLMs (\textit{MobFusion-C}), as graph connectors for fusing text and earth-observation embeddings (\textit{MobFusion-G}), and as graph tokens for MLLMs (\textit{MobFusion-T}), guided by urban sociological insights.

\item Experiments on large-scale datasets from three U.S. metropolitan areas, i.e., Chicago, Boston, and New York City, show that incorporating mobility consistently improves foundation models on socioeconomic prediction tasks, including median household income, population density, and crime prediction.
\end{itemize}

\section{Related Works}

\noindent \textbf{Foundation Models for Urban Geospatial Tasks.}
Large foundation models, including LLMs, MLLMs, and geospatial foundation models such as AlphaEarth, have been increasingly explored for geospatial understanding. GeoLLM~\cite{manvi2023geollm} prompts geographic coordinates together with auxiliary geographic context to predict geospatial indicators, such as population density and economic livelihoods. Its follow-up study~\cite{manvi2024large} further examines spatial biases in LLM-based geospatial inference. Beyond language models, geospatial foundation models~\cite{agarwal2024general,brown2025alphaearth,muhlematter2025urbanfusion} have also emerged. AlphaEarth Foundations~\cite{brown2025alphaearth} integrates large-scale multimodal Earth observation data into unified geospatial embeddings. 
Recent works use VLMs and MLLMs for urban socioeconomic prediction~\cite{liucitylens,liu2025cityrise,hao2025urbanvlp}, taking POI text, geospatial descriptions, satellite imagery, and street-view images as input, but still focus on location-level intrinsic attributes and lack mobility-based urban connectivity.

\noindent \textbf{Mobility in Computational Social Science.}
Urban mobility has been widely used to reflect urban socioeconomic conditions~\cite{barbosa2021uncovering,cagney2020urban,aiken2022machine,moro2021mobility}. Existing studies connect mobility networks to social segregation and exposure~\cite{athey2021estimating,moro2021mobility}, neighborhood income and economic performance~\cite{alves2021commuting,yoshimura2022revisiting}, crime~\cite{levy2020triple}, and pandemic vulnerability~\cite{chang2021mobility}. 
Beyond outcome prediction, another line of work analyzes the structural properties of mobility systems, including recurring motifs~\cite{schneider2013unravelling}, community organization~\cite{louail2015uncovering}, entropy~\cite{marin2022uncovering}, and commuting connectivity~\cite{alves2021commuting, li2026quantifying}.

\noindent \textbf{GNN-based Urban Mobility Modeling.} Human mobility has become a powerful tool for studying urban socioeconomic phenomena. Existing GNN-based methods~\cite{zhou2023heterogeneous,zhang2021multi,hui2020predicting} typically construct mobility networks with regions or POIs as nodes and mobility flows as edges, and learn node embeddings through self-supervised learning for downstream tasks. Some studies~\cite{wen2026mora} further treat mobility networks as a backbone or modality for multimodal urban representation learning, deriving region embeddings from mobility graphs and aligning them with visual or textual features.

\section{Preliminary and Problem Statement}

\begin{definition}[Mobility Network]
A \emph{mobility network} of a city is a CBG-POI heterogeneous graph
$
\mathcal{G} \;=\; \bigl(\mathcal{C},\, \mathcal{P},\, \mathcal{E}_{v},\, \mathcal{E}_{b}),
$
where $\mathcal{C} = \{c_{1}, \dots, c_{n_{c}}\}$ is the set of Census Block Group (CBG) nodes
and $\mathcal{P} = \{p_{1}, \dots, p_{n_{p}}\}$ is the set of point-of-interest (POI) nodes.
Two typed edge sets connect them:
the \emph{visit edges} $\mathcal{E}_{v} \subseteq \mathcal{C} \times \mathcal{P}$,
where $(c_{i}, p_{j}) \in \mathcal{E}_{v}$ if residents of $c_{i}$ make at least one recorded visit to $p_{j}$,
each carrying a weight $w_{ij} = \log\bigl(1 + n_{ij}\bigr)$ derived from the visit count $n_{ij}$ aggregated over the observation window;
and the \emph{belonging edges} $\mathcal{E}_{b} \subseteq \mathcal{P} \times \mathcal{C}$,
where $(p_{j}, c_{i}) \in \mathcal{E}_{b}$ if $p_{j}$ is physically located inside $c_{i}$.
\end{definition}

\noindent\textbf{Problem Statement:} 
For each CBG $c_{i}\in\mathcal{C}$, we observe optional static intrinsic
features $\mathbf{x}_{i}$ that may include satellite imagery, POI text, or
other geographic attributes, together with the mobility network $\mathcal{G}$
that provides relational context among CBGs and POIs.
Let $\boldsymbol{\Phi} = \{\Phi_{1}, \dots, \Phi_{K}\}$ denote a (possibly singleton) set of pre-trained
foundation models, e.g., LLMs, multimodal LLMs, or geospatial foundation models.
The task is to predict the socioeconomic indicator vector
$\mathbf{y}_{i} \in \mathbb{R}^{|\mathcal{Y}|}$
($\mathcal{Y}=\{\mathrm{income}, \mathrm{density}, \mathrm{crime}\}$) via a predictor
$f_{\theta}$ that composes (and optionally fine-tunes) those foundation models:
\begin{equation}
\widehat{\mathbf{y}}_{i} \;=\; f_{\theta} \bigl(\mathbf{x}_{i};\, \mathcal{G};\, \boldsymbol{\Phi}\bigr),
\label{eq:problem}
\end{equation}
where $\theta$ denotes the trainable parameters that adapt or fuse
$\boldsymbol{\Phi}$. 
This formulation is general: it covers zero-shot
prompting of a single foundation model as well as multi-model compositions
in which $\mathcal{G}$ is encoded by a dedicated graph module and fused with
the outputs of $\boldsymbol{\Phi}$. The central question is whether
incorporating $\mathcal{G}$ elicits geospatial-understanding capabilities of
$\boldsymbol{\Phi}$ beyond what static features $\mathbf{x}_{i}$ alone can
provide.

\section{Methodology}

In this section, we present \textit{MobFusion}, which consists of three variants: 
(i) \textit{MobFusion-C}, which treats mobility as contextual information for zero-shot LLM prediction; 
(ii) \textit{MobFusion-G}, which uses the mobility network as a graph connector to fuse visual and POI textual features of each CBG; 
and (iii) \textit{MobFusion-T}, which encodes the mobility network as graph tokens to enhance multimodal large language models.
\begin{figure*}[t!]
\centering
\begin{minipage}[t]{0.44\textwidth}
\begin{tcolorbox}[
  enhanced jigsaw,
  colback=gray!4, colframe=black!55,
  boxrule=0.5pt, arc=2pt,
  left=2mm, right=2mm, top=1mm, bottom=1mm,
  before skip=4pt, after skip=4pt,
  equal height group=prompt-cmp,
]
\setlength{\parskip}{0pt}\setlength{\parindent}{0pt}%
\textbf{\small  \texttt{(a) CBG Prompt only with POI features}}
\begin{Verbatim}[fontsize=\scriptsize, breaklines=true, baselinestretch=0.92]
You are analyzing a Census Block Group (CBG)
in Boston. **Location**:  42.3331 N, 71.1035 W
This area has 282 recorded places.

**Sample POI inside this area**:
 - Sully's Barber Shop -- Personal Care / Hair
 - Apts at 32 Worthington St -- Real Estate
 - Dunkin' -- Restaurants / Snack Bars
 - 7-Eleven -- Grocery / Convenience ... 

**Category distribution**:
 - Lessors of Real Estate: 171 (61%)
 - Offices of Physicians:   25  (9%)
 - Restaurants & Eating:    23  (8%) ... 

**Summary**: area dominated by Lessors of Real
Estate, also featuring Offices of Physicians
and Restaurants.

Based on what is physically located within this
area, estimate the socioeconomic characteristics
of this neighborhood.

Estimate (income/density/crime) on a 0.0--9.9 scale:

\end{Verbatim}
\end{tcolorbox}
\end{minipage}
\hfill
\begin{minipage}[t]{0.54\textwidth}
\begin{tcolorbox}[
  enhanced jigsaw,
  colback=blue!2, colframe=blue!55!black,
  boxrule=0.5pt, arc=2pt,
  left=2mm, right=2mm, top=1mm, bottom=1mm,
  before skip=4pt, after skip=4pt,
  equal height group=prompt-cmp,
]
\setlength{\parskip}{0pt}\setlength{\parindent}{0pt}%
\textbf{\sffamily\small\color{blue!55!black}  \texttt{(b) CBG prompts with additional mobility context}}
\begin{Verbatim}[fontsize=\scriptsize, breaklines=true, baselinestretch=0.92,commandchars=\\\{\}]
\textbf{Same as (a): Location, POI sample, Category distribution, Summary}

**Resident mobility profile** - where people living here travel:
 - Outflow: 234,784 trips (297 distinct dests)
 - Inflow:  1.7x city-avg visitor volume
 - 24% stay local; 76% travel elsewhere
 Most distinctive activities (vs city avg):
 - Video Tape & Disc Rental    (z=5.3,  1.3%)
 - Specialty Food Stores       (z=4.4,  1.3%)

**Consume locally**:    
Full-Service Restaurants, Specialty Food, ...

**Travel outside for**: 
Offices of Physicians, Fitness Centers, ...

**Attract visitors**:   
Full-Service Restaurants, Pharmacies, ...

Based on what is physically located within this area AND where 
residents actually travel, estimate the socioeconomic 
characteristics of this neighborhood.

Estimate (income/density/crime) on a 0.0--9.9 scale:

\end{Verbatim}
\end{tcolorbox}
\end{minipage}
\caption{Zero-shot prompt templates: (a) prompt with only intrinsic POI features (sampled POI names and categories) of the CBG; (b) builds on (a) by additionally incorporating the CBG's mobility neighborhood effects, i.e., internal, inflow, and outflow patterns.}
\label{fig:prompt-template}
\end{figure*}

\subsection{Mobility as Contexts for Zero-shot Geospatial Prediction with LLMs}
Unlike existing works~\cite{manvi2024large,manvi2023geollm} that convert a region’s coordinates, address information, and POI information into prompts, urban mobility data are usually represented as mobility networks and are difficult to prompt directly. Existing graph prompting methods~\cite{perozzi2024letgraphtalkingencoding,fatemi2024talk} also face scalability challenges on large mobility networks.
Inspired by Triple Neighborhood Effects~\cite{levy2020triple}, we summarize mobility networks from three perspectives: internal regional flows, inbound patterns, and outbound patterns. We then incorporate these summaries into prompts to enable efficient mobility-enhanced zero-shot prediction.

\begin{definition}[Triple Neighborhood Effects]

Inspired by the \emph{triple neighborhood disadvantage} perspective~\cite{levy2020triple}, we generalize the focus from disadvantage to broader neighborhood effects. 
For each CBG \(i\), we define \emph{triple neighborhood effects} through three CBG--POI mobility channels. 
(i) The \textbf{internal effect} captures visits made by residents of CBG \(i\) to POIs located within the same CBG, reflecting local activity intensity. 
(ii) The \textbf{outbound effect} captures visits made by residents of CBG \(i\) to POIs located in other CBGs, reflecting the external opportunities and activities. 
(iii) The \textbf{inbound effect} captures visits to POIs located in CBG \(i\) made by residents from other CBGs, reflecting how CBG \(i\) attracts external visitors.
\end{definition}

\noindent\textbf{Mobility Network as Context for Prompting a CBG:}
Figure~\ref{fig:prompt-template}(a) shows the basic CBG prompt, which only uses the intrinsic POI distribution and sampled POI names. 
This prompt describes what is physically located inside a CBG, but misses the mobility patterns of the CBG. 
We argue that the socioeconomic character of a CBG is more clearly revealed by where its residents \emph{travel}, who \emph{visit} it, and what \emph{circulates locally}, rather than by the supply-side POI list alone. 
We therefore extend the basic prompt with a structured mobility profile derived from the CBG-POI mobility network~$\mathcal{G}$. 

As shown in Figure~\ref{fig:prompt-template}(b), the mobility-enhanced prompt summarizes three types of information: 
(1) \emph{mobility statistics}, including total outflow trips, distinct destinations, inflow ratio over the city average, and internal/outflow share; 
(2) \emph{distinctive resident activities}, which identify POI categories that are unusually frequent for residents of the target CBG; 
and (3) \emph{triple neighborhood effects summary}, which separately describes what residents consume locally, what they travel outside for, and what attracts outside visitors to this CBG. 
For LLM outputs, following GeoLLM~\cite{manvi2023geollm}, we formulate prediction as a classification task by scaling each label value to the range from 0.0 to 9.9 and rounding it to one decimal place. 
This constrained label space makes LLM outputs easier to parse and reduces instability caused by free-form numerical generation.

\subsection{Mobility as Graph Connections for Multimodal City Embedding Fusion}

In this section, we examine whether the CBG-POI mobility network can serve as a relational connector for multimodal urban representation learning. 
\textit{MobFusion-G} propagates information between CBG-level visual embeddings and POI-level textual embeddings through the mobility graph for mobility-aware embedding fusion.

\paragraph{CBG–POI mobility network:}
In the mobility network $ \mathcal{G}$, for CBG node $c_i \in \mathcal{C}$,  its feature \(\mathbf{x}^{\mathrm{cbg}}_i\) is derived from one of two vision-based representations:
The first uses satellite images sampled within the CBG boundary, which are encoded by a frozen vision foundation model (e.g., RemoteCLIP~\cite{liu2024remoteclipvisionlanguagefoundation}) and then pooled into a CBG-level representation. 
The second uses AlphaEarth embeddings~\cite{brown2025alphaearth}, where 10m-resolution embeddings are extracted within the CBG polygon and pooled as the CBG feature. 

For each POI \(p_j\in\mathcal{P}\), we construct a text description (detailed in Table~\ref{tab:poi_feature_sample}) combining its name, category, and location, and encode it with a frozen text embedding model (e.g., BGE-M3~\cite{bge_m3}) to obtain its feature \(\mathbf{x}^{\mathrm{poi}}_j\).


\begin{figure}[h]
    \centering
    \includegraphics[width=1\linewidth]{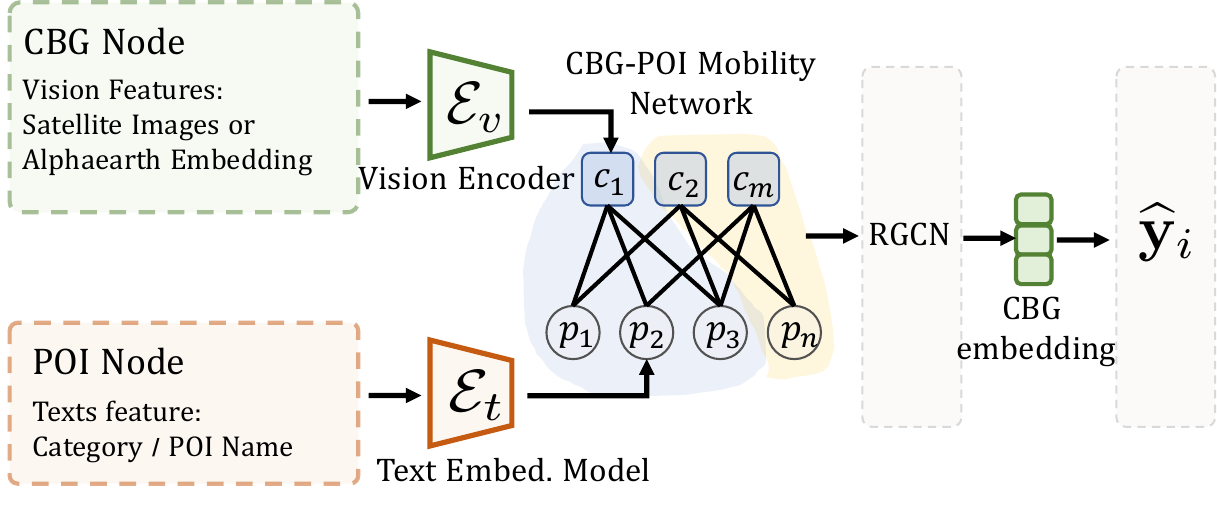}
    \caption{CBG-POI mobility network as the connector for the vision and text embedding fusion.}
    \label{fig:mobfusion-connector}
\end{figure}

For feature fusion and message passing, as shown in Figure~\ref{fig:mobfusion-connector}, we use a two-hop heterogeneous R-GCN~\cite{schlichtkrull2017modelingrelationaldatagraph} that respects the bipartite CBG-POI structure: each hop updates only one side of the graph. 
The relation set is \(\mathcal{R}=\{r_v,r_b\}\), where \(r_v\) denotes the weighted visit relation from CBGs to POIs, and \(r_b\) denotes the structural belonging relation from POIs to their host CBGs. 
Since CBG visual features and POI textual features have different dimensions, we first project them into a shared hidden space using
$
\mathbf{h}^{(0)}_{c_i}=\mathbf{W}_{c}\mathbf{x}^{\mathrm{cbg}}_{i}
$
and 
$
\mathbf{h}^{(0)}_{p_j}=\mathbf{W}_{p}\mathbf{x}^{\mathrm{poi}}_{j}
$.
The first hop updates each POI by aggregating messages from CBGs whose residents visit it (via \(r_v\)) and from the CBG that physically hosts it (via the reverse of \(r_b\)). 
The second hop updates each CBG by aggregating messages from POIs its residents visit (via the reverse of \(r_v\)), and from POIs it hosts (via \(r_b\)). 
This two-hop propagation realizes mobility-driven metapaths such as \(c_i \xrightarrow{r_v} p_j \xrightarrow{r_v^{-1}} c_k\) (CBGs connected through shared visit destinations) and \(c_i \xrightarrow{r_v} p_j \xrightarrow{r_b} c_k\) (CBGs connected via the physical locations of visited POIs).
The final output is a mobility-aware CBG embedding:
\begin{equation}
\mathbf{z}_i = \mathrm{R\text{-}GCN}\bigl(c_i;\mathcal{G},\{\mathbf{x}^{\mathrm{cbg}}_i\},\{\mathbf{x}^{\mathrm{poi}}_j\}\bigr)
\in \mathbb{R}^{d}.
\end{equation}

\paragraph{Self-supervised pretraining.}
We first pretrain the heterogeneous encoder without using socioeconomic labels. 
We create two stochastic views of the CBG--POI graph by randomly dropping edges and masking node features. 
The same R-GCN encoder maps each view into CBG representations, which are then passed through a projection head. 
For each CBG, the representations from the two views form a positive pair, while other CBGs in the batch serve as negative samples. 
Following SimCLR~\cite{chen2020simple}, we optimize a symmetric InfoNCE objective~\cite{oord2018representation} that pulls the two views of the same CBG together while pushing apart other CBGs in the batch.

\paragraph{Supervised fine-tuning.}
After pretraining, we attach a lightweight regression head \(h_{\psi}\) (a two-layer MLP) to the CBG embedding \(\mathbf{z}_i\) and predict \(\hat{\mathbf{y}}_i = h_{\psi}(\mathbf{z}_i)\). 
To balance the scale across heterogeneous tasks, we standardize labels per training fold to \(z\)-scores. 
The encoder and regression head are optimized jointly using mean squared error:
\begin{equation}
\mathcal{L}^{\mathrm{G}}_{\mathrm{sup}}
=
\frac{1}{|\mathcal{Y}|}
\sum_{t\in\mathcal{Y}}
\frac{1}{|\mathcal{C}^{\mathrm{train}}_t|}
\sum_{i\in\mathcal{C}^{\mathrm{train}}_t}
\bigl(\hat{y}_{i,t}-\tilde{y}_{i,t}\bigr)^2,
\end{equation}
where \(\tilde{y}_{i,t}\) denotes the standardized value for task $t$ and \(\mathcal{C}^{\mathrm{train}}_t\) is the set of training CBGs with valid labels for prediction task \(t\). 
At evaluation, predictions are inverse-transformed back to the raw scale.

\begin{figure}[h]
    \centering
    \includegraphics[width=\linewidth]{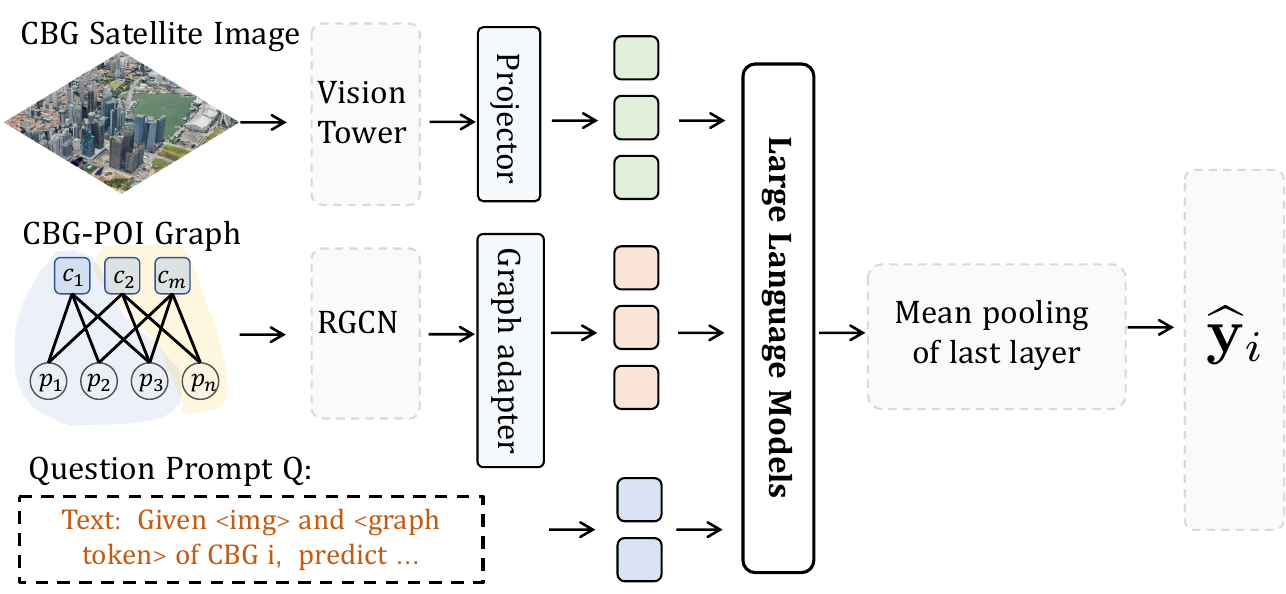}
    \caption{Mobility network as graph token for MLLM prediction.}
    \label{fig:mobfusion-token}
\end{figure}

\subsection{Mobility Network as Tokens: Enhance the Geospatial Understanding of MLLMs}

\paragraph{Mobility Graph Tokens.}
We further introduce \textit{MobFusion-T}, which injects mobility-aware graph representations into multimodal large language models as virtual graph tokens. 
For each CBG \(c_i\), we use the CBG embedding \(\mathbf{z}_i\in\mathbb{R}^{d}\) (\(d{=}128\)) produced by the \emph{self-supervised} stage of \textit{MobFusion-G} (before supervised fine-tuning), so that no socioeconomic labels are used to construct the graph representation. 
A lightweight graph adapter \(A_{\eta}\) (a two-layer MLP) maps \(\mathbf{z}_i\) into \(K\) token embeddings with the same hidden size as the language model:
\begin{equation}
\mathbf{G}_i = A_{\eta}(\mathbf{z}_i) \in \mathbb{R}^{K\times H},
\end{equation}
where \(H\) is the hidden dimension of the MLLM and we use \(K{=}4\) in our experiments. 
These graph tokens provide a compact representation of the CBG's mobility context and allow the mobility signal to participate in multimodal reasoning.

\paragraph{Mobility-enhanced Multimodal Inference and Fine-tuning.}
As shown in Figure~\ref{fig:mobfusion-token}, the MLLM input is structured as a sequence of satellite tiles, \(K\) graph tokens, and the question text, 
where the satellite tiles are sampled from the target CBG, the \(K\) graph tokens are placeholders for mobility information, and the question is either a \emph{basic} or \emph{POI-enriched} prompt. 
Following LLaVA-style multimodal injection~\cite{liu2023llava}, we introduce a dedicated graph token \(\langle\texttt{graph}\rangle\) in the input sequence and assign it the embedding \(\mathbf{G}_i\) before the sequence is consumed by the language model. 
This lifts mobility signals into the language embedding space alongside image and text tokens, allowing them to participate in self-attention across all transformer layers.

For urban prediction, we mean-pool the final hidden representations from the last transformer layer and pass them through a regression head:
\begin{equation}
\hat{\mathbf{y}}_i = h_{\psi}\!\left(
\mathrm{Pool}\bigl(\Phi(\mathbf{I}_i, \mathbf{G}_i, Q)\bigr)
\right),
\end{equation}
where \(\Phi\) denotes the multimodal language model and \(h_{\psi}\) denotes the two-layer MLP regression head.

During training, we keep the vision tower and the base language model weights frozen, and update only the graph adapter, LoRA modules~\cite{hu2022lora}, and the regression head. 
We optimize the trainable modules through:
\begin{equation}
\mathcal{L}^{\mathrm{T}}_{\mathrm{sup}}
=
\frac{1}{|\mathcal{C}_{\mathrm{train}}||\mathcal{Y}|}
\sum_{i\in\mathcal{C}_{\mathrm{train}}}
\sum_{t\in\mathcal{Y}}
\bigl(\hat{y}_{i,t}-\tilde{y}_{i,t}\bigr)^2,
\end{equation}
where \(\tilde{y}_{i,t}\) denotes the standardized value of each task. 
This design allows the MLLM to leverage mobility networks for geospatial prediction while requiring only lightweight adaptation.

\section{Evaluation}

\subsection{Datasets and Tasks}

We conduct experiments on three large-scale datasets covering Boston, Chicago, and New York City. 
The datasets are as follows. 

\begin{itemize}[leftmargin=*]
    \item \textbf{Mobility networks.} We use SafeGraph Monthly Patterns~\cite{safegraph_patterns} and Advan Neighborhood Patterns~\cite{advan_neighborhood_patterns} to construct CBG-POI mobility networks, where edges are weighted by aggregated visit volumes.
    \item \textbf{POI information.} We obtain POI attributes from SafeGraph Global Places~\cite{safegraph_global_places}, including POI names, categories, coordinates, and identifiers. POI text embeddings are produced by BGE-M3~\cite{bge_m3}.
    \item \textbf{Satellite imagery.} We collect the satellite imagery from the National Agriculture Imagery Program (NAIP)\footnote{\url{https://naip-usdaonline.hub.arcgis.com/}} and associate image tiles with CBGs through spatial intersection.
    \item \textbf{AlphaEarth embeddings.} We use AlphaEarth embeddings~\cite{brown2025alphaearth}, which provide 64-dimensional foundation-model representations at 10-meter resolution, and pool them within each CBG boundary.
\end{itemize}

\paragraph{Socioeconomic Groundtruth.}
We define three CBG-level urban prediction tasks, including (i) median household income prediction, (ii) population density prediction, and (iii) crime prediction. 
Ground-truth labels for income and population are obtained from the census data of 2023 released by the United States Census Bureau~\cite{census_tables}. 
Crime counts are derived from incident records maintained by the police departments of Chicago, New York City, and Boston.
Detailed data statistics and descriptions are in Appendix~\ref{app:dataset}.

\begin{table*}[t]
    \centering
    \small
    \setlength{\tabcolsep}{6pt}
    \renewcommand{\arraystretch}{1.2}
    \resizebox{0.8\textwidth}{!}{%
    \begin{tabular}{l l ccc ccc ccc}
    \toprule
    &
    & \multicolumn{3}{c}{\textbf{Boston}}
    & \multicolumn{3}{c}{\textbf{Chicago}}
    & \multicolumn{3}{c}{\textbf{NYC}} \\
    \cmidrule(lr){3-5} \cmidrule(lr){6-8} \cmidrule(lr){9-11}
    \textbf{Model} & \textbf{Variant}
    & Inc. & Den. & Cri. & Inc. & Den. & Cri. & Inc. & Den. & Cri. \\
    \midrule
    \multirow{2}{*}{GPT-4.1}
    & POI only         & 0.295 & 0.768 & 0.427 & 0.505 & 0.595 & 0.647 & 0.392 & 0.518 & \textbf{0.682} \\
    & Mobility-aware   & 0.349 & \textbf{0.774} & 0.440 & 0.555 & 0.609 & \textbf{0.658} & 0.402 & 0.537 & 0.672 \\
    \cmidrule(l){2-11}
    \multirow{2}{*}{Gemini-2.5-Flash}
    & POI only         & 0.238 & 0.651 & 0.329 & 0.461 & 0.492 & 0.551 & 0.369 & 0.430 & 0.655 \\
    & Mobility-aware   & 0.281 & 0.653 & 0.344 & 0.486 & 0.471 & 0.508 & 0.362 & 0.478 & 0.623 \\
    \cmidrule(l){2-11}
    \multirow{2}{*}{GPT-5.4}
    & POI only         & 0.360 & 0.765 & 0.435 & 0.539 & \textbf{0.640} & 0.560 & 0.408 & 0.582 & 0.668 \\
    & Mobility-aware   & \textbf{0.417} & 0.763 & \textbf{0.444} & \textbf{0.558} & 0.636 & 0.563 & \textbf{0.473} & \textbf{0.596} & 0.657 \\
    \bottomrule
    \end{tabular}%
    }
    \caption{Spearman~$\rho$ for median household income (Inc.), population density (Den.), and crime (Cri.) prediction across three cities under two prompt variants. Best per column in \textbf{bold}.}
    \label{tab:approach1}
    \end{table*}

\subsection{Experimental Settings}
\label{sec:exp-settings}

\paragraph{Metrics.} 

Following GeoLLM~\cite{manvi2024large,manvi2023geollm}, we use two metrics, 
Pearson's $r^2$ and Spearman's rank correlation $\rho$. 
The squared Pearson coefficient $r^2$ is commonly used in geospatial prediction tasks~\cite{manvi2023geollm,perez2017poverty,jean2016combining}, which measures the absolute fit on the original label value scale. 
Spearman's $\rho$ measures the rank consistency between predicted and ground-truth values. For task $t \in \{\text{income}, \text{density}, 
\text{crime}\}$,
we have $
\rho_t \;=\; \frac{\mathrm{Cov}\bigl(\mathrm{R}(\hat{y}_t),\, \mathrm{R}(y_t)\bigr)}
{\sigma_{\mathrm{R}(\hat{y}_t)} \, \sigma_{\mathrm{R}(y_t)}}, $
where $\hat{y}_t$ is the random variable of model-predicted scores for the task $t$, $y_t$ is the corresponding ground-truth random variable, 
$\mathrm{R}(\cdot)$ denotes the rank variable, and 
$\sigma_{\mathrm{R}(\cdot)}$ is its standard deviation. 
We choose Spearman's $\rho$ as our primary metric. The performance comparison on the squared Pearson  $r^2$ is shown in Appendix~\ref {app:metrics}.

\paragraph{Baselines and Implementation:} 
\begin{itemize}[leftmargin=*]
    \item For zero-shot prediction, we evaluate several LLM models, including GPT-4.1~\cite{achiam2023gpt}, Gemini-2.5-Flash~\cite{comanici2025gemini}, and GPT-5.4~\cite{openai2025gpt5}, under the same prompt setting. 
    \item For mobility as the connector evaluation, we compare our \textit{MobFusion-G} with ridge regression (RidgeCV)~\cite{hoerl1970ridge} and MORA~\cite{wen2026mora}. 
    RidgeCV uses: (i) satellite-image (encoded by RemoteCLIP~\cite{liu2024remoteclipvisionlanguagefoundation}), (ii) AlphaEarth embeddings, (iii) mean-pooled POI text embeddings, and (iv) the concatenation of POI embeddings with either satellite-image or AlphaEarth~\cite{brown2025alphaearth} embeddings as inputs. MORA~\cite{wen2026mora} aligns pretrained mobility-network embeddings with visual (AlphaEarth or satellite image embeddings) and POI representations through contrastive learning.
    \item For mobility as graph token evaluation,  we evaluate component ablations of the mobility-enhanced VLM (i.e., Qwen2.5-VL-7B\footnote{\url{https://qwenlm.github.io/blog/qwen2.5-vl/}}). 
    We vary the graph-token source, image input, and prompt content, and remove the graph token to test whether mobility information contributes beyond visual and textual inputs.
\end{itemize}
The detailed hyperparameters and settings are introduced in Appendix~\ref{app:Hyperparameters}.

\subsection{Zero-shot Prompting Performance}

Table~\ref{tab:approach1} reports the zero-shot prediction results of three LLMs under POI-only and mobility-aware prompt settings. 
Mobility context improves income prediction most consistently across three cities, with
gains up to +0.065 in Spearman~$\rho$ (GPT-5.4 on NYC).
However, its effects on population density and crime prediction are more mixed, with several settings showing marginal or negative changes. 
This heterogeneity is expected because population density is closely related to built-environment features, whereas reported crime can be affected by city-specific reporting and patterns. 
Notably, GPT-5.4 achieves the strongest absolute performance and the largest mobility-driven gains across most settings, suggesting that stronger reasoning capability amplifies the
utility of mobility context.

\begin{figure}[h]
    \centering
    \includegraphics[width=0.9\linewidth]{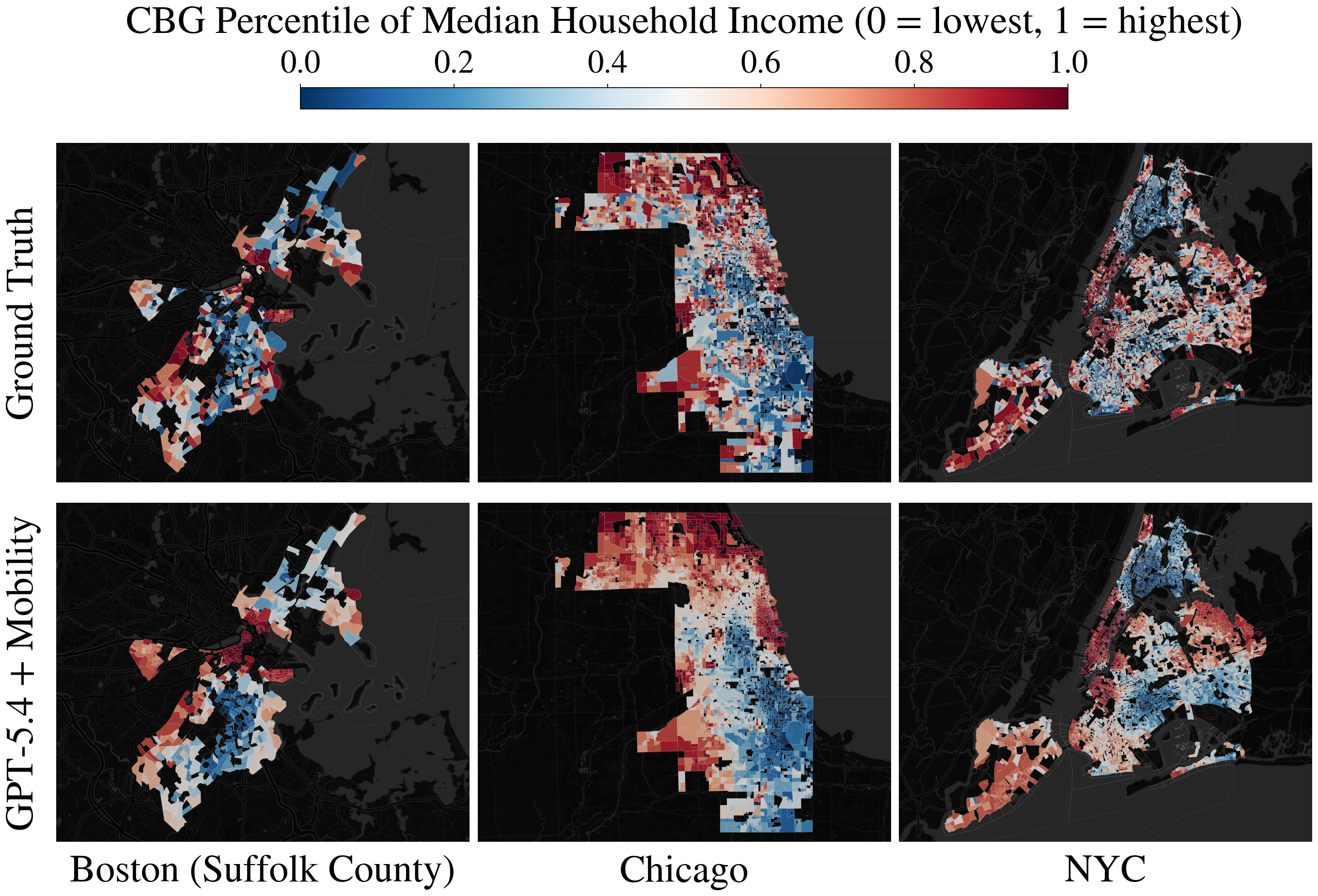}
    \caption{Spatial distribution of median household income percentile. Top: ground truth; bottom: zero-shot predictions from GPT-5.4 with mobility contexts.}
    \label{fig:LLM-zero-shot-visualization-income}
\end{figure}

Figure~\ref{fig:LLM-zero-shot-visualization-income} compares the spatial distribution of median household income percentile between the ground truth and the zero-shot prediction by GPT-5.4 with mobility contexts. 
We find that the zero-shot LLM exhibits non-trivial geospatial understanding: its predicted distribution matches the ground truth at the macro-regional scale across all three cities, correctly localizing broad high- and low-income areas. Yet it tends to over-smooth locally, assigning similar values to neighboring CBGs and missing the fine-grained income heterogeneity within. The spatial visualization of the other metrics is shown in Appendix~\ref{app:Spatial Visualization}.

\subsection{Mobility as Connector Evaluation}

\begin{table*}[t]
    \centering
    \small
    \setlength{\tabcolsep}{3pt}
    \renewcommand{\arraystretch}{1.15}
    \resizebox{\textwidth}{!}{%
    \begin{tabular}{l l ccc ccc ccc}
    \toprule
    &
    & \multicolumn{3}{c}{\textbf{Boston}}
    & \multicolumn{3}{c}{\textbf{Chicago}}
    & \multicolumn{3}{c}{\textbf{NYC}} \\
    \cmidrule(lr){3-5} \cmidrule(lr){6-8} \cmidrule(lr){9-11}
    \textbf{Method}                                & \textbf{Variant}
    & Inc. & Den. & Cri. & Inc. & Den. & Cri. & Inc. & Den. & Cri. \\
    \midrule
    \multirow{3}{*}{RidgeCV}
    & AE only                          & $0.587_{\pm.026}$ & $\underline{0.874}_{\pm.010}$ & $0.538_{\pm.025}$ & $0.611_{\pm.016}$ & $\cellcolor{blue!10}\mathbf{0.809}_{\pm.006}$ & $0.724_{\pm.017}$ & $0.570_{\pm.020}$ & $\cellcolor{blue!10}\mathbf{0.774}_{\pm.030}$ & $0.461_{\pm.026}$ \\
    & POI only                         & $0.432_{\pm.043}$ & $0.741_{\pm.022}$ & $\cellcolor{blue!10}\mathbf{0.662}_{\pm.020}$ & $0.518_{\pm.045}$ & $0.591_{\pm.042}$ & $0.831_{\pm.004}$ & $0.470_{\pm.023}$ & $0.570_{\pm.028}$ & $0.348_{\pm.035}$ \\    
    & AE $\Vert$ POI                   & $0.563_{\pm.009}$ & $0.863_{\pm.017}$ & $0.583_{\pm.034}$ & $0.591_{\pm.026}$ & $0.742_{\pm.018}$ & $0.830_{\pm.023}$ & $0.563_{\pm.032}$ & $\underline{0.759}_{\pm.020}$ & $0.476_{\pm.018}$ \\
    & RemoteCLIP $\Vert$ POI           & $0.581_{\pm.043}$ & $0.781_{\pm.019}$ & $0.559_{\pm.031}$ & $0.615_{\pm.005}$ & $0.658_{\pm.025}$ & $0.829_{\pm.008}$ & $0.601_{\pm.022}$ & $0.670_{\pm.026}$ & $0.459_{\pm.031}$ \\
    \cmidrule(l){2-11}
    \multirow{2}{*}{MORA}
    & AE $\Vert$ POI        & $0.490_{\pm.027}$ & $0.850_{\pm.013}$ & $0.576_{\pm.017}$ & $0.578_{\pm.026}$ & $0.714_{\pm.009}$ & $0.861_{\pm.008}$ & $0.529_{\pm.026}$ & $0.703_{\pm.021}$ & $0.511_{\pm.025}$ \\
    & RemoteCLIP $\Vert$ POI           & $\underline{0.613}_{\pm.023}$ & $0.867_{\pm.012}$ & $0.612_{\pm.014}$ & $0.648_{\pm.008}$ & $0.773_{\pm.019}$ & $0.876_{\pm.008}$ & $\underline{0.625}_{\pm.026}$ & $0.721_{\pm.016}$ & ${0.526}_{\pm.031}$ \\
    \cmidrule(l){2-11}
    \multirow{2}{*}{MobFusion-G}
    & AE                    & $0.600_{\pm.019}$ & $\cellcolor{blue!10}\mathbf{0.885}_{\pm.013}$ & $\underline{0.644}_{\pm.012}$ & $\underline{0.650}_{\pm.018}$ & $\underline{0.804}_{\pm.015}$ & $\underline{0.872}_{\pm.006}$ & $0.588_{\pm.026}$ & $0.757_{\pm.016}$ & $\underline{\mathbf{0.530}}_{\pm.016}$ \\
    & RemoteCLIP                       & $\cellcolor{blue!10}\mathbf{0.616}_{\pm.026}$ & $0.861_{\pm.014}$ & $0.610_{\pm.028}$ & $\cellcolor{blue!10}\mathbf{0.661}_{\pm.016}$ & $0.770_{\pm.015}$ & \cellcolor{blue!10}$\mathbf{0.878}_{\pm.013}$ & $\cellcolor{blue!10}\mathbf{0.627}_{\pm.021}$ & $0.727_{\pm.012}$ & $\cellcolor{blue!10}\mathbf{0.530}_{\pm.029}$ \\
    \bottomrule
    \end{tabular}%
    }
    \caption{Five-fold results of Spearman~$\rho$.
    Three method families with two visual encoders for CBGs (i.e., AlphaEarth or embedding of satellite image). Blue color marks the best per column; underline marks the second best.}
    \label{tab:approach2}
\end{table*}
Table~\ref{tab:approach2} shows whether the CBG-POI mobility network can serve as a connector for multimodal embedding fusion. 
Across the three cities, \textit{MobFusion-G} achieves the strongest results on income prediction and also performs best on crime prediction in Chicago and New York City. 
The gains on population density are less consistent, as AlphaEarth embeddings already provide strong built-environment signals for this task. 
Overall, the results show that mobility-based relational propagation adds useful information beyond intrinsic visual and POI representations, especially for income and safety-related prediction.

\begin{figure}[h]
    \centering
    \includegraphics[width=1\linewidth]{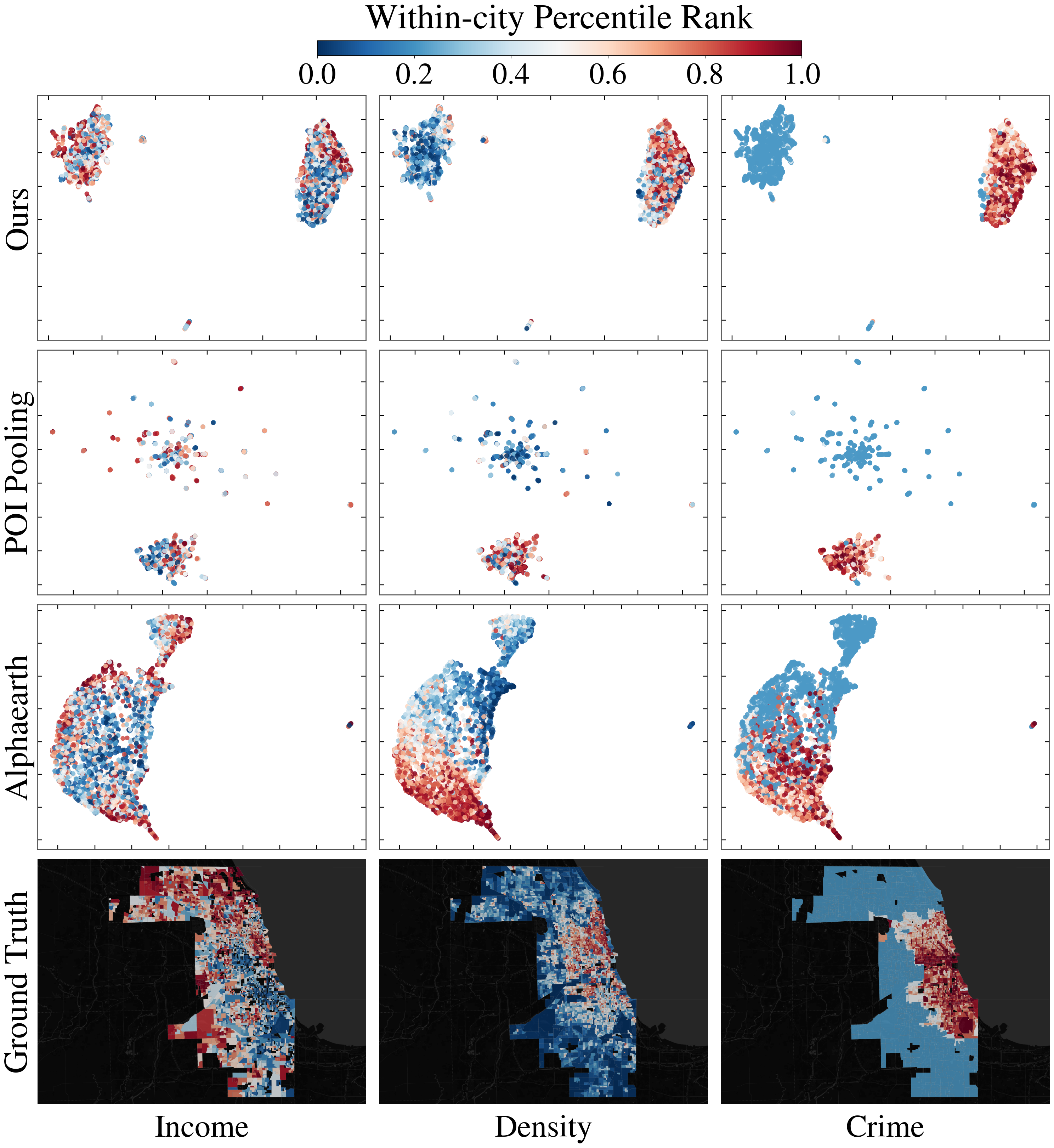}
    \caption{UMAP visualization of CBG embeddings in Chicago, colored by percentile rank of income, density, and crime (columns).}
    \label{fig:fusion-visualization-chicago}
\end{figure}

Figure~\ref{fig:fusion-visualization-chicago} shows UMAP~\cite{mcinnes2018umap} projections of Chicago CBG embeddings colored by income, density, and crime percentile. The embeddings of \textit{MobFusion-G} display clearer socioeconomic stratification than POI-only and AlphaEarth embeddings, with high- and low-percentile CBGs forming more coherent clusters, which indicates that mobility-enhanced fusion yields a latent space better aligned with urban socioeconomic structure.

\subsection{Mobility-aware MLLM Performance}

To evaluate whether mobility graph tokens enhance MLLMs, we compare several
input variants of \textit{MobFusion-T} along three dimensions: the visual
feature used to construct the graph token, the visual input, and the textual
prompt. 
For the graph token, CBG nodes in the mobility network are initialized
with either AlphaEarth or RemoteCLIP embeddings and
serialized into tokens fed to the MLLM. 
The visual input consists of one or four satellite tiles sampled within the
target CBG boundary.
The textual prompt is either \emph{basic} (city name plus the
task question) or \emph{POI-enriched} (basic prompt augmented with a
textualized description of the POI distribution within the CBG).
\begin{table}[t]
    \centering
    \small
    \setlength{\tabcolsep}{4pt}
    \renewcommand{\arraystretch}{1.25}
    \resizebox{\columnwidth}{!}{%
    \begin{tabular}{lllccc}
    \toprule
    \textbf{Graph token} & \textbf{Prompt} & \textbf{Image} & \textbf{Inc.} & \textbf{Den.} & \textbf{Cri.} \\
    \midrule
    AlphaEarth & POI-enriched & 1 tile  & $0.611_{\pm.017}$ & $\cellcolor{blue!10}\mathbf{0.778}_{\pm.013}$ & $\cellcolor{blue!10}\mathbf{0.709}_{\pm.031}$ \\
    AlphaEarth & POI-enriched & 4 tiles & $\cellcolor{blue!10}\mathbf{0.614}_{\pm.017}$ & $\underline{0.776}_{\pm.019}$ & $\underline{0.704}_{\pm.022}$ \\
    RemoteCLIP & POI-enriched & 1 tile  & $0.599_{\pm.018}$ & $0.728_{\pm.009}$ & $0.689_{\pm.031}$ \\
    RemoteCLIP & POI-enriched & 4 tiles & $0.604_{\pm.017}$ & $0.744_{\pm.015}$ & $0.694_{\pm.029}$ \\
    \midrule
    AlphaEarth & Basic & 4 tiles & $\underline{0.613}_{\pm.023}$ & $0.775_{\pm.009}$ & $0.694_{\pm.027}$ \\
    AlphaEarth & POI-enriched & \textemdash & $0.569_{\pm.011}$ & $0.763_{\pm.014}$ & $0.702_{\pm.030}$ \\
    \textemdash & Basic & 4 tiles & $0.605_{\pm.016}$ & $0.732_{\pm.018}$ & $0.635_{\pm.025}$ \\
    \bottomrule
    \end{tabular}}
\caption{Ablation results of mobility-enhanced MLLM variants. Blue marks the best per column; underline marks the second best. The top block contrasts graph tokens and image inputs under the POI-enriched prompt. The bottom block ablates one component at a time: the prompt, the image input, or the graph token.}
    \label{tab:03_mllm_ablation}
\end{table}

Table~\ref{tab:03_mllm_ablation} reports the ablation results on median household income, population density, and crime prediction. 
The full setting, which combines mobility graph tokens, satellite images, and POI-enriched prompts, achieves the best overall performance. 
Removing the graph token causes clear drops, especially on population density and crime, showing that mobility structure provides useful information beyond visual and textual inputs. 
The remaining ablations suggest that image inputs and POI-enriched prompts contribute complementary signals for geospatial prediction.


\section{Conclusion}

In this work, we examine whether human mobility can enhance foundation models’ understanding of urban socioeconomic conditions by complementing intrinsic visual and textual features. We propose a mobility-enhanced foundation-modeling paradigm for urban tasks, inspired by modular AI. 
Experiments show that mobility networks improve LLMs, fusion of geospatial foundation models, and MLLMs on downstream urban tasks. These findings can encourage broader NLP research on urban mobility and smart cities.

\section*{Limitations}

Our framework operates at the Census Block Group (CBG) level, which
reflects an inherent trade-off between label fidelity and behavioural
resolution. Finer units (e.g.,  Census Block) carry substantially
noisier ACS estimates and are often suppressed for privacy; coarser
units (Census Tract, ZIP code) smooth out the very mobility signal we
exploit, collapsing the CBG-POI visit graph to near-uniform aggregate
flows. CBGs are the smallest unit at which both the ACS labels remain
trustworthy, and the SafeGraph mobility edges retain meaningful spatial
structure. A finer-grained study would require an alternative ground
truth (administrative or sensor-derived) that we leave to future work.

\paragraph{Use of AI assistants.}
We used Claude, ChatGPT, and Gemini for language polishing and code assistance during this work. All scientific contributions, including ideas, experiments, and analyses, are the authors' own.

\section*{Ethical Considerations}

This work raises no individual-level privacy concerns. All ground-truth labels come from publicly released ACS 2023 estimates and city
open-data crime portals, both already aggregated with privacy
thresholds. The mobility data are SafeGraph monthly and neighborhood
patterns, aggregated to CBG level by the provider; we access no
individual trajectories, no device-level location traces, and make no
attempt to re-identify users. Satellite imagery is pre-licensed
daytime tiles containing no personally identifiable information.

\bibliography{custom}

\appendix

\section{Appendix}
\label{sec:appendix}



\subsection{Details of Datasets}
\label{app:dataset}

\begin{table}[h]
\centering
\small
\setlength{\tabcolsep}{8pt}
\renewcommand{\arraystretch}{1.2}
\resizebox{\columnwidth}{!}{%
\begin{tabular}{lrrr}
\toprule
 & \textbf{Boston} & \textbf{Chicago} & \textbf{NYC} \\
\midrule
CBGs with ACS labels  & 3{,}273 & 4{,}002 & 6{,}807 \\
POIs visited volume   & 136{,}284 & 151{,}972 & 190{,}052 \\
CBG-POI edges         & 44{,}751{,}964 & 87{,}565{,}413 & 31{,}745{,}979 \\
Total CBG-POI visits  & 9.5\,$\times 10^{8}$ & 1.6\,$\times 10^{9}$ & 3.5\,$\times 10^{8}$ \\
\bottomrule
\end{tabular}%
}
\caption{Per-city statistics of the mobility data.}
\label{tab:dataset_stats}
\end{table}

\begin{table}[t]
    \centering
    \small
    \setlength{\tabcolsep}{6pt}
    \renewcommand{\arraystretch}{1.2}
    \resizebox{\columnwidth}{!}{%
    \begin{tabular}{ll}
    \toprule
    \textbf{Field} & \textbf{Value} \\
    \midrule
    \texttt{PLACEKEY}        & \texttt{zzy-227@62j-sg9-3t9} \\
    \texttt{LOCATION\_NAME}  & Ria Money Transfer Partner Location \\
    \texttt{TOP\_CATEGORY}   & Activities Related to Credit Intermediation \\
    \texttt{SUB\_CATEGORY}   & Financial Transactions Processing, Reserve, \\
                             & and Clearinghouse Activities \\
    \texttt{NAICS\_CODE}     & 522320 \\
    \texttt{LATITUDE}        & 42.305241 \\
    \texttt{LONGITUDE}       & -71.114442 \\
    \texttt{STREET\_ADDRESS} & 140A South St \\
    \texttt{CITY}            & Jamaica Plain \\
    \texttt{REGION}          & MA \\
    \texttt{POSTAL\_CODE}    & 02130 \\
    \texttt{cbg\_geoid}      & 250251202011 \\
    \midrule
    \multicolumn{2}{p{0.92\columnwidth}}{\textbf{Generated poi description for Text embedding model}: \par\smallskip
    \emph{Location name of the POI is Ria Money Transfer Partner Location.
    POI address is 140A South St, Jamaica Plain. Approximate coordinates:
    $(42.30524, -71.11444)$. TOP\_CATEGORY: Activities Related to Credit
    Intermediation. SUB\_CATEGORY: Financial Transactions Processing,
    Reserve, and Clearinghouse Activities.}} \\
    \bottomrule
    \end{tabular}%
    }
    \caption{Example POI feature record and the generated text description of a POI.}
    \label{tab:poi_feature_sample}
\end{table}

\paragraph{Urban Mobility Data.} 
We utilize  SafeGraph Monthly Patterns~\cite{safegraph_patterns} and Advan Neighborhood Patterns dataset~\cite{advan_neighborhood_patterns} to construct mobility networks for three cities. Both datasets aggregate footfall data at the census block group (CBG) level over a one-month period, capturing how populations move between areas and places in terms of both volume and frequency. For each city, we build a mobility network in which nodes are CBGs and POIs, and edges are weighted by visit volume.

\paragraph{Points of Interest (POI).} 
We obtain the Points of Interest data of three cities from the SafeGraph Global Places dataset~\cite{safegraph_global_places}.
The POI attributes consist of the ID, NAICS-style category information, textual POI name, and coordinates. 
Per-POI text embeddings are produced by BGE-m3~\cite{bge_m3}  for use as POI node
features in our mobility graph.
The POI data samples are introduced in Table~\ref{tab:poi_feature_sample}.

\paragraph{Satellite Imagery.} 
The satellite imagery is obtained from the high-resolution National Agriculture Imagery Program (NAIP)\footnote{\url{https://naip-usdaonline.hub.arcgis.com/}} for the three study cities. We organize the imagery into a $1\mathrm{km} \times 1\mathrm{km}$ grid and store each grid cell as a $512 \times 512$ RGB tile. For each CBG, we identify overlapping image tiles through spatial intersection, use the nearest tile as a fallback when no overlap exists, and establish the corresponding CBG–image associations.

\paragraph{AlphaEarth Embedding.} 
The AlphaEarth embedding dataset~\cite{brown2025alphaearth} provides global 64-dimensional foundation-model representations for every $10\mathrm{m} \times 10\mathrm{m}$ location on Earth. We use these embeddings to characterize the geospatial environment of each CBG and mean-pool all corresponding 10 m pixel-level vectors into a compact region-level representation. 

\paragraph{Socioeconomic Groundtruth.} We introduce the socioeconomic data sources in three large cities, Boston, Chicago, and New York City, as follows:
\begin{itemize}[leftmargin=*]
    \item \textbf{Median household income} is taken from the American Community Survey 5-Year Estimates (ACS) 2023 release~\cite{census_tables}. Values are inflation-adjusted U.S.\ dollars at the CBG resolution.
  \item \textbf{Population density}:  For each CBG~$c$ we compute
  $\text{density}_{c} = \text{pop}_{c} \,/\, \text{area}_{c}$
  in residents per km$^2$. The numerator $\text{pop}_{c}$ is the total
  population from the ACS 2023 table. The denominator $\text{area}_{c}$ is the geodesic area of the CBG polygon.
  \item \textbf{Crime Count}: We collect 2023 incident-level records from each city's public safety open-data portal: 
the Boston Police Department~\cite{boston_crime_2023}, 
the City of Chicago Data Portal~\cite{chicago_crime_2023}, 
and NYC OpenData ``NYPD Complaint Data Historic''~\cite{nyc_crime_historic}.
\end{itemize}

\subsection{Detailed Metric Computation}
\label{app:metrics}

Building on the brief description in the main text, this section
provides the full formulae and explains how the two metrics are
computed for each of the three settings.

  \paragraph{Spearman's rank correlation $\rho$.}
  Spearman's $\rho$ measures the rank consistency between predicted and
  ground-truth values. Let $\hat{y}_t$ denote the random variable of
  model predictions for task $t \in {\text{income}, \text{density},
  \text{crime}}$, $y_t$ the corresponding ground-truth random variable,
  $\mathrm{R}(\cdot)$ the rank-transform operator, and
  $\sigma{\mathrm{R}(\cdot)}$ the standard deviation of the ranks. The
  Spearman coefficient is
  \begin{equation}
  \rho_t \;=\; \frac{\mathrm{Cov}\bigl(\mathrm{R}(\hat{y}_t),, \mathrm{R}(y_t)\bigr)}
  {\sigma{\mathrm{R}(\hat{y}_t)} , \sigma{\mathrm{R}(y_t)}}.
  \end{equation}
  Because $\rho_t$ depends only on the ranks of the predictions, it is
  \emph{invariant under any strictly monotone transform} applied to
  either side. This invariance is what lets us place three settings with
  incommensurable prediction scales onto a single comparable axis (see
  below).

  \paragraph{Pearson's $r^2$.}
  The squared Pearson correlation
  \begin{equation}
  r^2_t \;=\; \left(\frac{\mathrm{Cov}(\hat{y}_t,, y_t)}
  {\sigma{\hat{y}_t}, \sigma{y_t}}\right)^{2}
  \end{equation}
  captures the linear fit on the original label scale and is sensitive to
  both scale and bias. Following GeoLLM~\cite{manvi2023geollm} and the
  broader satellite-based socio-economic prediction
  literature~\cite{perez2017poverty,jean2016combining}, we report
  $r^2_t$ alongside $\rho_t$ as a secondary metric whenever the
  prediction is numerically comparable to the label value on its native scale.

\paragraph{Per-setting Metric Calculation.}
The three settings produce predictions on different scales, which
defined as follows:
\begin{itemize}[leftmargin=*, itemsep=2pt]
\item \textbf{\textit{MobFusion-C} (Zero-shot LLM prediction).} 
Following GeoLLM~\cite{manvi2023geollm}, the frozen LLM is prompted to output a single score \(\hat{y}_t \in [0.0, 9.9]\) for each CBG and task. 
We scale the ground-truth task values to the same range before evaluation, so the predicted scores can be compared with the scaled labels. 
\item \textbf{\textit{MobFusion-G} (Mobility as graph connector ).} The mobility graph encoder is followed by a
regression head that predicts different tasks in a $z$-scored space, where
$\mu_t, \sigma_t$ are estimated on the training-fold labels. 
At evaluation time, we inverse-transform the prediction
$\hat{y}_t = \sigma_t \hat{z}_t + \mu_t$. Both
$\rho_t$ and $r^2_t$ are then computed on this raw scale.
\item \textbf{\textit{MobFusion-T} (Mobility as graph tokens for MLLM).} 
The model produces per-task predictions in a \(z\)-scored space, where the normalization is computed per city and per train fold. 
For single-city evaluation, we invert the city-specific \(z\)-score and compute \(r_t^2\) on the original scale of different tasks. 
For joint evaluation across Boston, Chicago, and New York City, we report city-normalized Spearman's \(\rho_t\). 
Specifically, predictions and task labels are standardized within each city before concatenation, so the joint \(\rho_t\) measures within-city ranking ability rather than being dominated by inter-city scale differences.
\end{itemize}

\paragraph{K-fold aggregation.}
All metrics are computed on five-fold experiments. We report the per-fold mean $\pm$ standard deviation across the five folds.

\subsection{Hyperparameters.}
\label{app:Hyperparameters}
For \textbf{\textit{MobFusion-G} (Mobility as graph connector)}, we
employ a two-layer relational graph convolutional network  over
the CBG-POI graph. CBG nodes are
initialised with either a 64-dimensional AlphaEarth embedding or a
768-dimensional RemoteCLIP embedding, while POI nodes are
initialised with the 1024-dimensional textual embedding; all
hidden representations are projected to 128 dimensions. We first pre-train the encoder for 300 epochs with the InfoNCE objective
($\tau = 0.07$), and then fine-tune it for an additional 150 epochs
under a supervised regression loss with a learning rate of
$1 \times 10^{-3}$.

For \textbf{\textit{MobFusion-T} (Mobility as graph tokens for MLLM)},
we keep the Qwen2.5-VL-7B backbone fully frozen and inject LoRA
adapters (rank $r = 16$) into the $q$ and $v$ projections of all 28
transformer layers. A lightweight GraphAdapter projects the 128-d CBG
embedding produced by \textit{MobFusion-G} to $K \times 3584$ graph
tokens ($K = 4$), which are concatenated with the text and image
tokens at the input. A linear regression head ($3584 \to 3$) is
attached on top of the last hidden state to produce the three task
predictions. The trainable parameters are optimized with AdamW (learning rate $2 \times 10^{-5}$) for 10 epochs. 

All training and inference are conducted on four NVIDIA L40S GPUs with 44 GB of memory. 


\subsection{More Experimental Results}

\noindent \textbf{(1) Pearson's $r^2$ metric for Zero-shot LLM Prediction Experiments:} Table~\ref{tab:approach1_r2} shows the Pearson's $r^2$ results comparison between different LLM models across two prompt settings (i.e., prompts with POI information, and prompts with mobility-aware contexts). From Table~\ref{tab:approach1_r2}, we find that the results are broadly consistent with the Spearman results in Table~\ref {tab:approach1}. Mobility-aware prompts generally improve income and population-density prediction over POI-only prompts, especially for GPT-5.4. The gains on crime prediction are more mixed, suggesting that crime is harder to infer from prompt-level mobility summaries alone. 

\noindent \textbf{(2) Pearson's $r^2$ metric for Mobility as Connector evaluation:} 
Table~\ref{tab:approach2_r2} reports the Pearson's \(r^2\) results for the mobility-as-connector setting. 
The results are consistent with the Spearman results in the main text. 
\textit{MobFusion-G} achieves the best performance on income prediction across all three cities, showing that CBG-POI mobility propagation is effective for capturing socioeconomic variation. 
For crime prediction, \textit{MobFusion-G} obtains the best or second-best results in most cities. 
For population density, the gains are less consistent, since vision-only features, especially AlphaEarth embeddings, already provide strong cues about the built environment and population concentration.

\noindent \textbf{(3) Per-city Pearson's $r^2$ for Mobility-as-Graph-Token.}
Table~\ref{tab:approach3_r2} reports per-city $r^2$, complementing the joint
Spearman results in Table~\ref{tab:03_mllm_ablation}. 
First, the configuration, AlphaEarth graph token, POI-enriched prompt, and 4 image tiles, is the most consistent across cities and tasks, achieving the best or second-best $r^2$ on 7 of 9 (city, task) columns, which indicates that the three input modalities contribute complementary signals. 
Second, for income and density, all three inputs are needed, whereas for crime in Chicago and NYC, removing the satellite image yields the best $r^2$, suggesting that local visual cues introduce noise at this setting. 
Overall, these results support the value of mobility graph tokens while also showing that their benefit depends on the prediction task and city context.

\begin{table*}[h]
    \centering
    \small
    \setlength{\tabcolsep}{6pt}
    \renewcommand{\arraystretch}{1.2}
    \resizebox{0.85\textwidth}{!}{%
    \begin{tabular}{l l ccc ccc ccc}
    \toprule
    &
    & \multicolumn{3}{c}{\textbf{Boston}}
    & \multicolumn{3}{c}{\textbf{Chicago}}
    & \multicolumn{3}{c}{\textbf{NYC}} \\
    \cmidrule(lr){3-5} \cmidrule(lr){6-8} \cmidrule(lr){9-11}
    \textbf{Model} & \textbf{Variant}
    & Inc. & Den. & Cri. & Inc. & Den. & Cri. & Inc. & Den. & Cri. \\
    \midrule
    \multirow{2}{*}{GPT-4.1}
    & POI only          & 0.072 & 0.324 & 0.044 & 0.228 & 0.184 & 0.200 & 0.137 & 0.106 & \textbf{0.147} \\
    & Mobility-aware    & 0.098 & 0.366 & 0.056 & 0.275 & 0.203 & \textbf{0.232} & 0.150 & 0.136 & 0.142 \\
    \cmidrule(l){2-11}
    \multirow{2}{*}{Gemini-2.5-Flash}
    & POI only          & 0.036 & 0.250 & 0.032 & 0.172 & 0.156 & 0.164 & 0.097 & 0.071 & 0.119 \\
    & Mobility-aware    & 0.060 & 0.282 & 0.034 & 0.209 & 0.185 & 0.130 & 0.109 & 0.115 & 0.096 \\
    \cmidrule(l){2-11}
    \multirow{2}{*}{GPT-5.4}
    & POI only          & 0.126 & 0.357 & 0.055 & 0.259 & 0.258 & 0.158 & 0.112 & 0.175 & 0.126 \\
    & Mobility-aware    & \textbf{0.173} & \textbf{0.384} & \textbf{0.069} & \textbf{0.304} & \textbf{0.270} & 0.174 & \textbf{0.269} & \textbf{0.198} & 0.121 \\
    \bottomrule
    \end{tabular}%
    }
    \caption{Coefficient of Pearson's $r^2$  on the CBG-level prediction across all three cities. Each LLM emits scores in $[0, 9.9]$. \textbf{Inc.}/\textbf{Den.}/\textbf{Cri.} denote median household income, population density, and crime count. Best per column in \textbf{bold}.}
    \label{tab:approach1_r2}
\end{table*}

\begin{table*}[t]
    \centering
    \small
    \setlength{\tabcolsep}{3pt}
    \renewcommand{\arraystretch}{1.15}
    \resizebox{\textwidth}{!}{%
    \begin{tabular}{l l ccc ccc ccc}
    \toprule
    &
    & \multicolumn{3}{c}{\textbf{Boston}}
    & \multicolumn{3}{c}{\textbf{Chicago}}
    & \multicolumn{3}{c}{\textbf{NYC}} \\
    \cmidrule(lr){3-5} \cmidrule(lr){6-8} \cmidrule(lr){9-11}
    \textbf{Method}                                & \textbf{Variant}
    & Inc. & Den. & Cri. & Inc. & Den. & Cri. & Inc. & Den. & Cri. \\
    \midrule
    \multirow{4}{*}{RidgeCV}
    & POI only                       & $0.139_{\pm.069}$ & $0.304_{\pm.053}$ & $0.078_{\pm.033}$ & $0.235_{\pm.067}$ & $0.223_{\pm.044}$ & $0.371_{\pm.063}$ & $0.200_{\pm.028}$ & $0.269_{\pm.038}$ & $0.068_{\pm.018}$ \\
    & AE only                        & $0.352_{\pm.030}$ & $\cellcolor{blue!10}\mathbf{0.633}_{\pm.09}$ & $0.094_{\pm.038}$ & $0.376_{\pm.025}$ & $\cellcolor{blue!10}\mathbf{0.577}_{\pm.071}$ & $0.370_{\pm.064}$ & $0.353_{\pm.015}$ & $\underline{0.485}_{\pm.022}$ & $0.178_{\pm.037}$ \\
    & AE $\Vert$ POI                 & $0.302_{\pm.033}$ & $0.616_{\pm.086}$ & $0.095_{\pm.027}$ & $0.332_{\pm.047}$ & $0.522_{\pm.068}$ & $0.411_{\pm.087}$ & $0.325_{\pm.043}$ & $0.477_{\pm.025}$ & $0.163_{\pm.061}$ \\
    & RemoteCLIP $\Vert$ POI         & $0.289_{\pm.049}$ & $0.476_{\pm.092}$ & $-0.033_{\pm.162}$ & $0.370_{\pm.024}$ & $0.369_{\pm.08}$ & $0.446_{\pm.077}$ & $0.360_{\pm.038}$ & $0.375_{\pm.04}$ & $0.164_{\pm.104}$ \\
    \cmidrule(l){2-11}
    \multirow{2}{*}{MORA}
    & AE $\Vert$ POI      & $0.236_{\pm.036}$ & $0.557_{\pm.074}$ & $0.115_{\pm.053}$ & $0.337_{\pm.024}$ & $0.440_{\pm.018}$ & $0.462_{\pm.118}$ & $0.301_{\pm.023}$ & $0.426_{\pm.03}$ & $0.187_{\pm.031}$ \\
    & RemoteCLIP $\Vert$ POI         & $0.338_{\pm.016}$ & $0.542_{\pm.087}$ & $\cellcolor{blue!10}\mathbf{0.188}_{\pm.107}$ & $0.431_{\pm.014}$ & $0.459_{\pm.094}$ & $0.521_{\pm.097}$ & $\cellcolor{blue!10}\mathbf{0.413}_{\pm.034}$ & $0.436_{\pm.026}$ & $\underline{0.242}_{\pm.073}$ \\
    \cmidrule(l){2-11}
    \multirow{2}{*}{MobFusion-G}
    & AE                  & $\underline{0.356}_{\pm.027}$ & $\underline{0.620}_{\pm.084}$ & $0.172_{\pm.061}$ & $\underline{0.434}_{\pm.02}$ & $\underline{0.564}_{\pm.026}$ & $\underline{0.504}_{\pm.11}$ & $\underline{0.372}_{\pm.026}$ & $\cellcolor{blue!10}\mathbf{0.488}_{\pm.028}$ & $0.212_{\pm.049}$ \\
    & RemoteCLIP                     & $\cellcolor{blue!10}\mathbf{0.370}_{\pm.031}$ & $0.557_{\pm.083}$ & $\underline{0.186}_{\pm.077}$ & $\cellcolor{blue!10}\mathbf{0.439}_{\pm.017}$ & $0.507_{\pm.03}$ & $\cellcolor{blue!10}\mathbf{0.532}_{\pm.105}$ & $\cellcolor{blue!10}\mathbf{0.413}_{\pm.03}$ & $0.441_{\pm.029}$ & $\cellcolor{blue!10}\mathbf{0.249}_{\pm.095}$ \\
    \bottomrule
    \end{tabular}%
    }
    \caption{\textbf{MobFusion-G evaluation.} 5-fold mean~$\pm$~std of Pearson's $r^2$.  Blue highlight marks the best, and underline marks the second best, in each (city, target) column.}
    \label{tab:approach2_r2}
\end{table*}

\begin{table*}[h]
    \centering
    \small
    \setlength{\tabcolsep}{4pt}
    \renewcommand{\arraystretch}{1.25}
    \resizebox{\textwidth}{!}{%
    \begin{tabular}{lll ccc ccc ccc}
    \toprule
    & & 
    & \multicolumn{3}{c}{\textbf{Boston}}
    & \multicolumn{3}{c}{\textbf{Chicago}}
    & \multicolumn{3}{c}{\textbf{NYC}} \\
    \cmidrule(lr){4-6} \cmidrule(lr){7-9} \cmidrule(lr){10-12}
    \textbf{Graph token} & \textbf{Prompt} & \textbf{Image}
    & Inc. & Den. & Cri. & Inc. & Den. & Cri. & Inc. & Den. & Cri. \\
    \midrule
    AlphaEarth & POI-enriched & 1 tile  & $0.168_{\pm.087}$ & $0.426_{\pm.101}$ & $0.130_{\pm.127}$ & $0.361_{\pm.031}$ & $0.511_{\pm.064}$ & $0.536_{\pm.102}$ & $0.374_{\pm.025}$ & $\cellcolor{blue!10}\mathbf{0.462}_{\pm.042}$ & $\underline{0.274}_{\pm.076}$ \\
    AlphaEarth & POI-enriched & 4 tiles  & $\underline{0.233}_{\pm.043}$ & $\underline{0.440}_{\pm.085}$ & $\cellcolor{blue!10}\mathbf{0.157}_{\pm.105}$ & $0.371_{\pm.044}$ & $\cellcolor{blue!10}\mathbf{0.557}_{\pm.066}$ & $\underline{0.550}_{\pm.111}$ & $\cellcolor{blue!10}\mathbf{0.393}_{\pm.035}$ & $\underline{0.457}_{\pm.037}$ & $0.242_{\pm.058}$ \\
    RemoteCLIP & POI-enriched & 1 tile  & $0.164_{\pm.074}$ & $0.245_{\pm.115}$ & $0.001_{\pm.244}$ & $0.366_{\pm.021}$ & $0.511_{\pm.115}$ & $0.521_{\pm.114}$ & $0.367_{\pm.053}$ & $0.401_{\pm.041}$ & $0.261_{\pm.086}$ \\
    RemoteCLIP & POI-enriched & 4 tiles  & $0.200_{\pm.044}$ & $0.263_{\pm.123}$ & $-0.034_{\pm.281}$ & $\underline{0.374}_{\pm.047}$ & $0.522_{\pm.078}$ & $0.497_{\pm.087}$ & $0.359_{\pm.03}$ & $0.401_{\pm.031}$ & $0.227_{\pm.099}$ \\
    \midrule
    AlphaEarth & Basic & 4 tiles  & $0.222_{\pm.043}$ & $\cellcolor{blue!10}\mathbf{0.441}_{\pm.073}$ & $0.120_{\pm.179}$ & $\cellcolor{blue!10}\mathbf{0.378}_{\pm.018}$ & $\underline{0.540}_{\pm.047}$ & $0.544_{\pm.098}$ & $0.388_{\pm.029}$ & $0.421_{\pm.075}$ & $0.237_{\pm.083}$ \\
    AlphaEarth & POI-enriched & \textemdash  & $0.153_{\pm.062}$ & $0.352_{\pm.128}$ & $-0.066_{\pm.186}$ & $0.354_{\pm.038}$ & $0.511_{\pm.063}$ & $\cellcolor{blue!10}\mathbf{0.587}_{\pm.122}$ & $0.319_{\pm.03}$ & $0.448_{\pm.038}$ & $\cellcolor{blue!10}\mathbf{0.294}_{\pm.1}$ \\
    \midrule
    \textemdash & Basic & 4 tiles  & $\cellcolor{blue!10}\mathbf{0.241}_{\pm.053}$ & $0.377_{\pm.07}$ & $\underline{0.153}_{\pm.094}$ & $0.338_{\pm.039}$ & $0.526_{\pm.065}$ & $0.425_{\pm.066}$ & $\underline{0.391}_{\pm.023}$ & $0.387_{\pm.037}$ & $0.151_{\pm.032}$ \\
    \bottomrule
    \end{tabular}%
    }
    \caption{\textbf{Per-city Pearson's $r^2$ (5-fold mean~$\pm$~std) for Mobility as Graph token evaluation.} Blue highlight marks the best, and underline marks the second best, in each (city, target) column.}
    \label{tab:approach3_r2}
\end{table*}

\subsection{Spatial Visualization}
\label{app:Spatial Visualization}

\begin{figure}[t]
    \centering
    \includegraphics[width=1\linewidth]{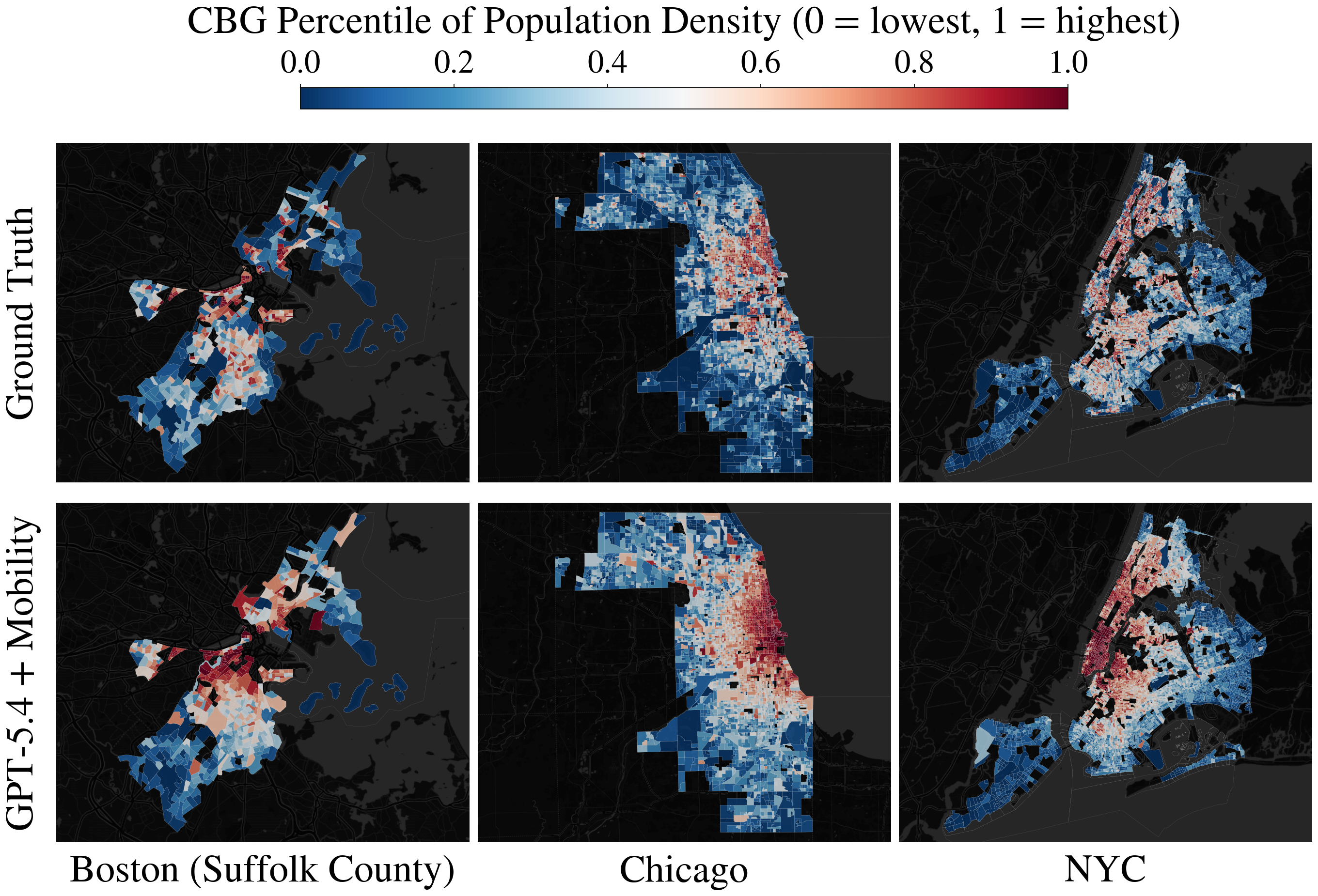}
    \caption{Spatial distribution of population density percentile. Top: ground truth; bottom: zero-shot predictions from GPT-5.4 with mobility contexts.}
    \label{fig:LLM-zero-shot-visualization-density}
\end{figure}

\begin{figure}[t]
    \centering
    \includegraphics[width=1\linewidth]{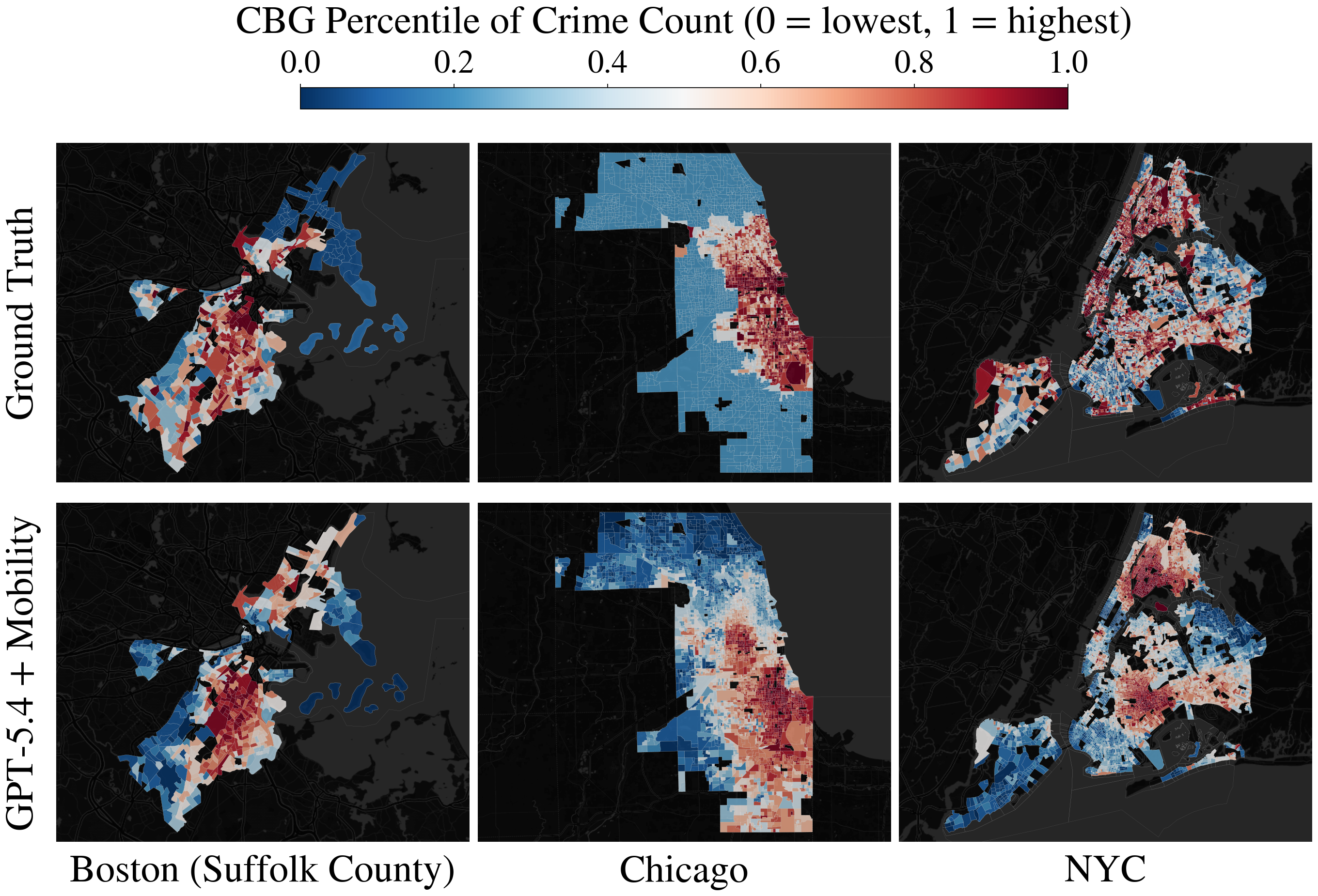}
    \caption{Spatial distribution of crime count percentile. Top: ground truth; bottom: zero-shot predictions from GPT-5.4 with mobility contexts.}
    \label{fig:LLM-zero-shot-visualization-crime}
\end{figure}

\noindent \textbf{Spatial Visualization of LLM Zero-shot Prediction:} Figure~\ref{fig:LLM-zero-shot-visualization-density} and Figure~\ref{fig:LLM-zero-shot-visualization-crime}  provide additional spatial visualizations for population density and crime count. 
For population density, GPT-5.4 with mobility contexts captures broad high-density areas in each city, especially the urban cores, but still smooths local variation across neighboring CBGs. For crime count, the predictions recover several city-level hotspots, but the spatial match is less stable than for population density.

\begin{figure}[h]
    \centering
    \includegraphics[width=1\linewidth]{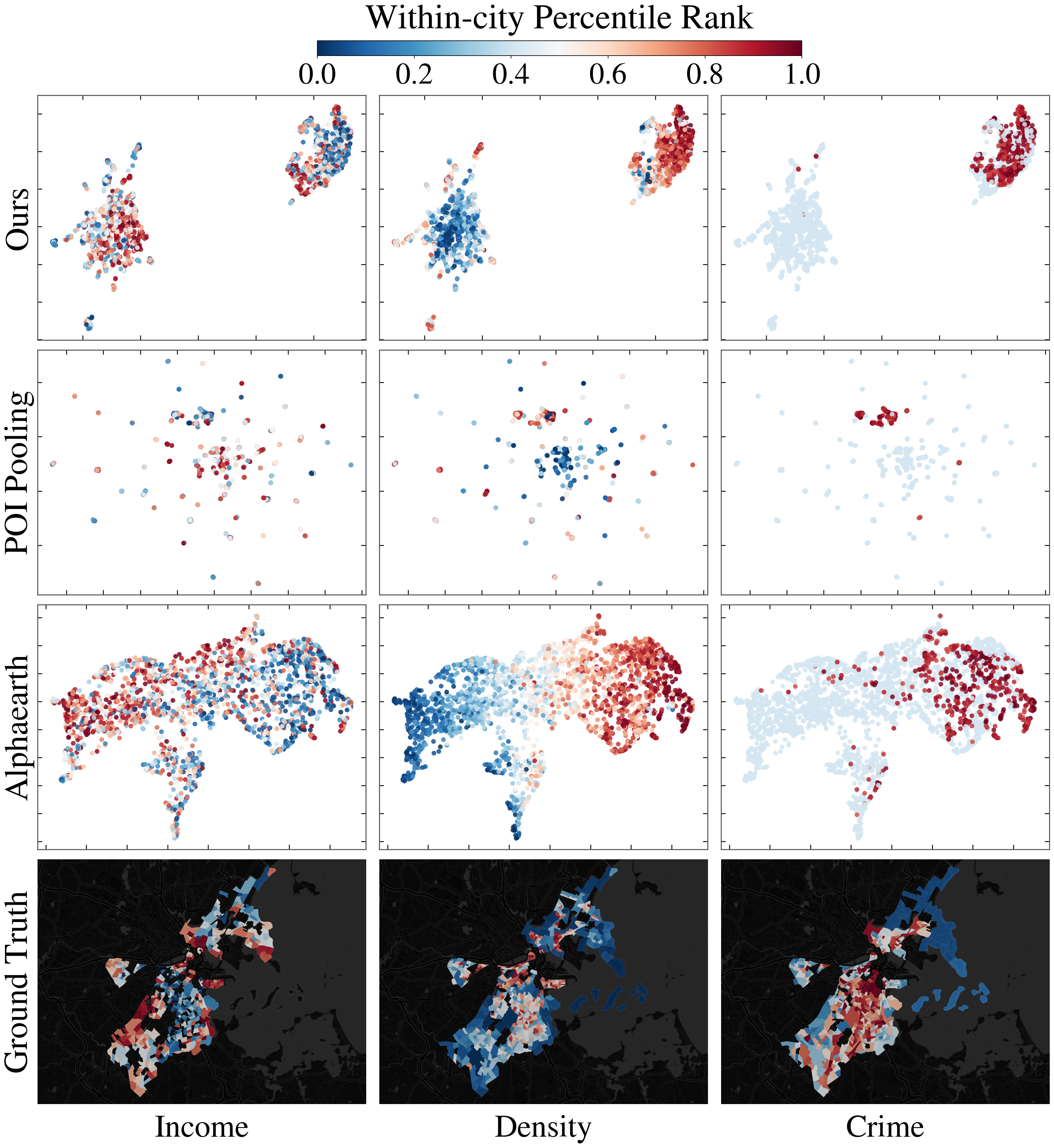}
    \caption{UMAP visualization of CBG embeddings in Boston, colored by percentile rank of income, density, and crime (columns). Rows compare AlphaEarth, POI embedding, and \textit{MobFusion-G} embedding.}
    \label{fig:embedding-visualization-boston}
\end{figure}

\begin{figure}[h]
    \centering
    \includegraphics[width=1\linewidth]{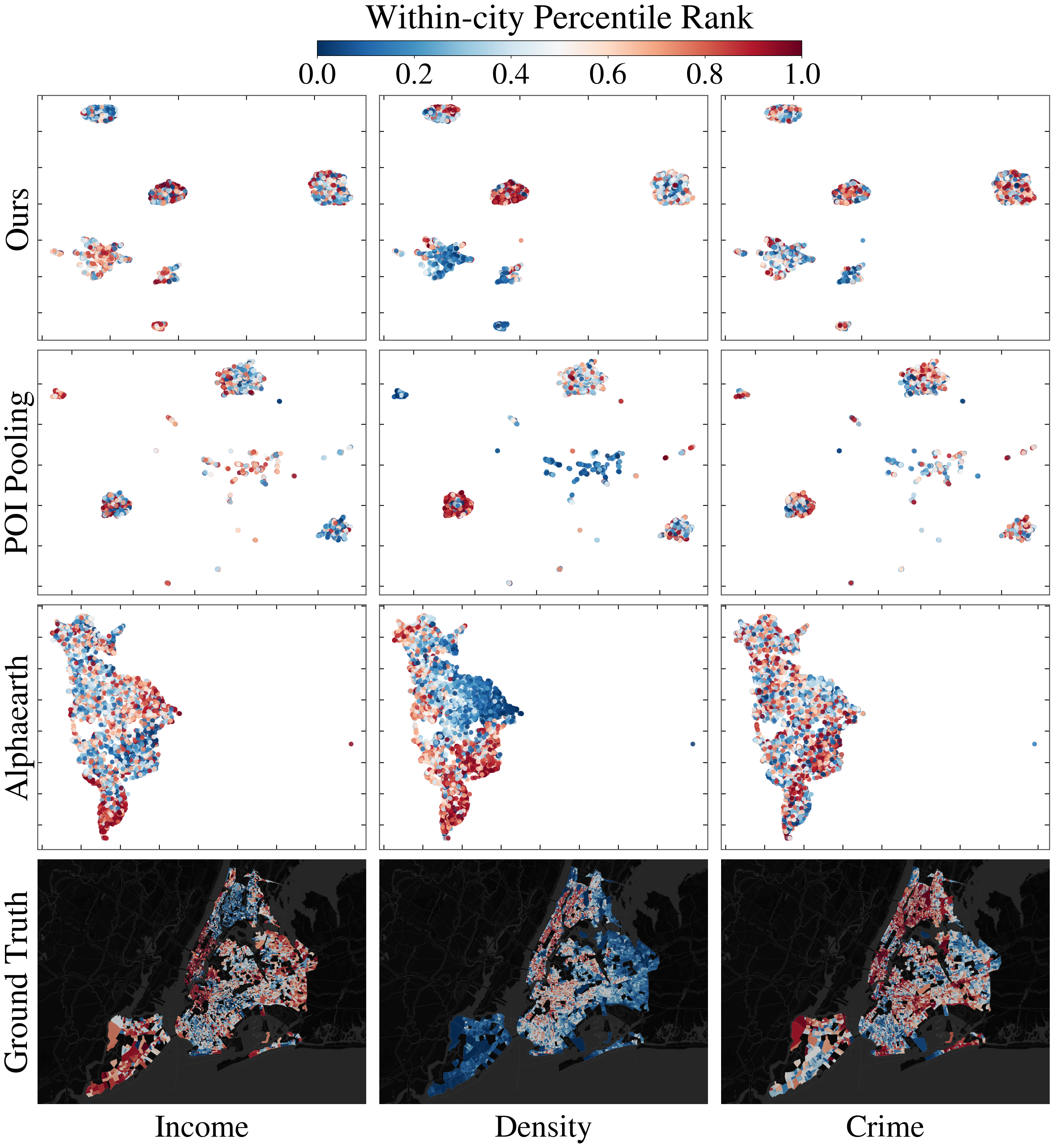}
    \caption{UMAP visualization of CBG embeddings in New York, colored by percentile rank of income, density, and crime (columns). Rows compare AlphaEarth, POI embedding, and \textit{MobFusion-G} embedding.}
    \label{fig:embedding-visualization-NYC}
\end{figure}

\noindent \textbf{UMAP Visualization of AlphaEarth embedding and Our \textit{MobFusion-G} Embedding:} Figure~\ref{fig:embedding-visualization-boston} and Figure~\ref{fig:embedding-visualization-NYC}  show UMAP visualization for Boston and New York City. Across both cities, \textit{MobFusion-G} embeddings show clearer separation between high- and low-percentile CBGs than POI-only embeddings, especially for income and population density. Compared with AlphaEarth embeddings, \textit{MobFusion-G} better preserves socioeconomic gradients in the latent space, suggesting that mobility-based relational fusion captures information beyond static built-environment features.

\subsection{Prompt examples}
\label{sec:appendix:prompts}
Fig.~\ref{fig:appendix-prompt-poi} and Fig.~\ref{fig:appendix-prompt-mobility} show example prompts that consider only the intrinsic POI features and the mobility contexts, respectively.

\begin{figure*}[t!]
\centering
\begin{tcolorbox}[
  enhanced jigsaw,
  colback=gray!4, colframe=black!55,
  boxrule=0.5pt, arc=2pt,
  left=2mm, right=2mm, top=1.5mm, bottom=1.5mm,
]
\setlength{\parskip}{0pt}\setlength{\parindent}{0pt}%
\textbf{\small Full prompt --- \texttt{CBG Prompt only with POI features}}
\begin{Verbatim}[fontsize=\scriptsize, breaklines=true, baselinestretch=0.95]
You are analyzing a Census Block Group (CBG) in Boston.

**Location**: approximately 42.3331 N, 71.1035 W

This area has 282 recorded places and businesses within its boundaries.

**Sample of businesses/locations inside this area** (8 shown):
  - Sully's Barber & Styling Shop -- Personal Care Services / Hair, Nail, and Skin Care Services
  - Apartments at 32 Worthington St -- Lessors of Real Estate / Lessors of Residential Buildings and Dwellings
  - Dunkin' -- Restaurants and Other Eating Places / Snack and Nonalcoholic Beverage Bars
  - Apartments at 70 Calumet St -- Lessors of Real Estate / Lessors of Residential Buildings and Dwellings
  - Apartments at 87 Hillside St -- Lessors of Real Estate / Lessors of Residential Buildings and Dwellings
  - Apartments at 62 Calumet St -- Lessors of Real Estate / Lessors of Residential Buildings and Dwellings
  - Apartments at 1560 Tremont St -- Lessors of Real Estate / Lessors of Residential Buildings and Dwellings
  - 7-Eleven -- Grocery Stores / Convenience Stores

**Category distribution** (how places are distributed by type):
  - Lessors of Real Estate: 171 place(s) (61%)
  - Offices of Physicians: 25 place(s) (9%)
  - Restaurants and Other Eating Places: 23 place(s) (8%)
  - Offices of Dentists: 8 place(s) (3%)
  - Offices of Real Estate Agents and Brokers: 6 place(s) (2%)

**Summary**: 
area dominated by Lessors of Real Estate, also featuring Offices of Physicians, Restaurants and Other Eating Places

Based on what is physically located within this area, estimate the socioeconomic characteristics of this neighborhood.

Please estimate the following on a scale from 0.0 to 9.9:
 - 0.0 = very low  (e.g. lowest income / lowest density / safest)
 - 9.9 = very high (e.g. highest income / highest density / most crime)

Respond ONLY with valid JSON: {"income": <float>, "density": <float>, "crime": <float>}
\end{Verbatim}
\end{tcolorbox}
\caption{Full POI-only prompt template instantiated on a
representative. No mobility information is
provided.}
\label{fig:appendix-prompt-poi}
\end{figure*}

\begin{figure*}[h]
\centering
\begin{tcolorbox}[
  enhanced jigsaw,
  colback=blue!2, colframe=blue!55!black,
  boxrule=0.5pt, arc=2pt,
  left=2mm, right=2mm, top=1.5mm, bottom=1.5mm,
]
\setlength{\parskip}{0pt}\setlength{\parindent}{0pt}%
\textbf{\small\color{blue!55!black} Full prompt --- \texttt{CBG Prompt with additional mobility context}}
\begin{Verbatim}[fontsize=\scriptsize, breaklines=true, baselinestretch=0.95]
You are analyzing a Census Block Group (CBG) in Boston.

**Location**: approximately 42.3331 N, 71.1035 W

This area has 282 recorded places and businesses within its boundaries.

**Sample of businesses/locations inside this area** (8 shown):
  - Sully's Barber & Styling Shop -- Personal Care Services / Hair, Nail, and Skin Care Services
  - Apartments at 32 Worthington St -- Lessors of Real Estate / Lessors of Residential Buildings and Dwellings
  - Dunkin' -- Restaurants and Other Eating Places / Snack and Nonalcoholic Beverage Bars
  - Apartments at 70 Calumet St -- Lessors of Real Estate / Lessors of Residential Buildings and Dwellings
  - Apartments at 87 Hillside St -- Lessors of Real Estate / Lessors of Residential Buildings and Dwellings
  - Apartments at 62 Calumet St -- Lessors of Real Estate / Lessors of Residential Buildings and Dwellings
  - Apartments at 1560 Tremont St -- Lessors of Real Estate / Lessors of Residential Buildings and Dwellings
  - 7-Eleven -- Grocery Stores / Convenience Stores

**Category distribution** (how places are distributed by type):
  - Lessors of Real Estate: 171 place(s) (61%)
  - Offices of Physicians: 25 place(s) (9%)
  - Restaurants and Other Eating Places: 23 place(s) (8%)
  - Offices of Dentists: 8 place(s) (3%)
  - Offices of Real Estate Agents and Brokers: 6 place(s) (2%)

**Summary**: area dominated by Lessors of Real Estate, also featuring Offices of Physicians, Restaurants and Other Eating Places

**Resident mobility profile** -- where people living here actually travel:
  - Outflow: 234,784 trips to other neighborhoods | 297 distinct destinations
  - Inflow:  1.7x city average visitor volume
  - 24% of resident trips stay local; 76% travel elsewhere

  Most distinctive activities of residents (vs city average):
    - Video Tape and Disc Rental (z=5.3, 1.3% of visits)
    - All Other Specialty Food Stores (z=4.4, 1.3% of visits)
    - Offices of Physicians (except Mental Health Specialists) (z=2.6, 36.9% of visits)

  What residents consume locally:
    - Full-Service Restaurants
    - Pharmacies and Drug Stores
    - All Other Specialty Food Stores

  What residents travel outside for:
    - Offices of Physicians (except Mental Health Specialists)
    - Fitness and Recreational Sports Centers
    - Full-Service Restaurants

  What attracts outside visitors here:
    - Full-Service Restaurants
    - Pharmacies and Drug Stores
    - All Other Specialty Food Stores

  Places residents frequently visit outside this area:
    - Dunkin'
    - Wentworth Institute Of Technology
    - Massachusetts College of Art and Design
    - Alexandra V Roth

Based on what is physically located within this area AND where residents actually travel, estimate the socioeconomic characteristics of this neighborhood.

Please estimate the following on a scale from 0.0 to 9.9:
 - 0.0 = very low  (e.g. lowest income / lowest density / safest)
 - 9.9 = very high (e.g. highest income / highest density / most crime)

Respond ONLY with valid JSON: {"income": <float>, "density": <float>, "crime": <float>}
\end{Verbatim}
\end{tcolorbox}
\caption{Full mobility-aware prompt template (\textit{MobFusion-C}) on the same
Boston CBG as Figure~\ref{fig:appendix-prompt-poi}. The header (location, POI
sample, category distribution, summary) is identical to the POI-only template;
the appended \emph{Resident mobility profile} block adds outflow/inflow
volumes, z-scored distinctive activities, and the three-flow consumption breakdown
(local consumption, outflow demand, inflow attraction).}
\label{fig:appendix-prompt-mobility}
\end{figure*}

\end{document}